\documentclass[conference]{IEEEtran}
\usepackage{cite}
\usepackage{amsmath,amssymb,amsfonts}
\usepackage{algorithm}
\usepackage{algorithmic}
\usepackage{graphicx}
\usepackage{textcomp}
\usepackage{xcolor}
\usepackage[hyphens]{url}
\usepackage{fancyhdr}
\usepackage{hyperref}
\usepackage{svg}
\usepackage{float}
\usepackage{subfig}
\usepackage{pifont}
\usepackage{multirow}
\usepackage[normalem]{ulem}
\usepackage{color}
\usepackage{bm}
\usepackage{cleveref}

\title{InstInfer: In-Storage Attention Offloading for Cost-Effective Long-Context LLM Inference}
\author{
	\IEEEauthorblockN{ \large
		Xiurui Pan\IEEEauthorrefmark{1}, 
		Endian Li\IEEEauthorrefmark{1}, 
            Qiao Li\IEEEauthorrefmark{2}, 
		Shengwen Liang\IEEEauthorrefmark{3}, 
		Yizhou Shan\IEEEauthorrefmark{4}, \\
            Ke Zhou\IEEEauthorrefmark{5},
            Yingwei Luo\IEEEauthorrefmark{1}, 
            Xiaolin Wang\IEEEauthorrefmark{1},
            and Jie Zhang\IEEEauthorrefmark{1}}

    \textit{\normalsize Computer Hardware and System Evolution Laboratory} \\

    {\normalsize Peking University\IEEEauthorrefmark{1}, \ Xiamen University\IEEEauthorrefmark{2}, \ Institute of Computing Technology, Chinese Academy of Sciences\IEEEauthorrefmark{3}} \\
    {\normalsize Huawei Cloud\IEEEauthorrefmark{4}, \  Wuhan National Laboratory for Optoelectronics of Huazhong University of Science and Technology\IEEEauthorrefmark{5}} \\
    {\normalsize https://www.chaselab.wiki}
}

\begin{document}
\maketitle

%%%%%%%%%%%%%%%%%%%%%%%%%%%%%%%%%%%%%
%%%%%%%%%% -- DO NOT MODIFY -- %%%%%%%%%%
%%%%%%%%%%%%%%%%%%%%%%%%%%%%%%%%%%%%%

% \author{
%   \ifdefined\hpcacameraready
%     \IEEEauthorblockN{\hpcaauthors{}}
%       \IEEEauthorblockA{
%         \hpcaaffiliation{} \\
%         \hpcaemail{}
%       }
%   \else
%     \IEEEauthorblockN{\normalsize{HPCA \hpcayear{} Submission
%       \textbf{\#\hpcasubmissionnumber{}}} \\
%       \IEEEauthorblockA{
%         Confidential Draft \\
%         Do NOT Distribute!!
%       }
%     }
%   \fi 
% }

% Heading and footer for title page
\fancypagestyle{camerareadyfirstpage}{%
  \fancyhead{}
  \renewcommand{\headrulewidth}{0pt}
  \fancyhead[C]{
    \ifdefined\aeopen
    \parbox[][12mm][t]{13.5cm}{\hpcayear{} IEEE International Symposium on High-Performance Computer Architecture (HPCA)}    
    \else
      \ifdefined\aereviewed
      \parbox[][12mm][t]{13.5cm}{\hpcayear{} IEEE International Symposium on High-Performance Computer Architecture (HPCA)}
      \else
      \ifdefined\aereproduced
      \parbox[][12mm][t]{13.5cm}{\hpcayear{} IEEE International Symposium on High-Performance Computer Architecture (HPCA)}
      \else
      \parbox[][0mm][t]{13.5cm}{\hpcayear{} IEEE International Symposium on High-Performance Computer Architecture (HPCA)}
    \fi 
    \fi 
    \fi 
    \ifdefined\aeopen 
      \includegraphics[width=12mm,height=12mm]{ae-badges/open-research-objects.pdf}
    \fi 
    \ifdefined\aereviewed
      \includegraphics[width=12mm,height=12mm]{ae-badges/research-objects-reviewed.pdf}
    \fi 
    \ifdefined\aereproduced
      \includegraphics[width=12mm,height=12mm]{ae-badges/results-reproduced.pdf}
    \fi
  }
  %\fancyfoot[L]{\hpcapubid{} \copyright \hpcayear{} IEEE}
  \fancyfoot[C]{}
}
% Heading and footer for remaining pages
\fancyhead{}
\renewcommand{\headrulewidth}{0pt}
%\fancyhead[C]{\hpcayear{} IEEE International Symposium on
% High-Performance Computer Architecture (HPCA)}

\ifdefined\hpcacameraready 
  \thispagestyle{camerareadyfirstpage}
  \pagestyle{empty}
\else
  \thispagestyle{plain}
  \pagestyle{plain}
\fi

\newcommand{\hpcaheight}{0mm}
\ifdefined\eaopen
\renewcommand{\hpcaheight}{12mm}
\fi

%Enables the camera ready header and footer

\begin{abstract}
The widespread of Large Language Models (LLMs) marks a significant milestone in generative AI. 
Nevertheless, the increasing context length and batch size in offline LLM inference escalate the memory requirement of the key-value (KV) cache, which imposes a huge burden on the GPU VRAM, especially for resource-constraint scenarios (e.g., edge computing and personal devices). Several cost-effective solutions leverage host memory or SSDs to reduce storage costs for offline inference scenarios and improve the throughput. Nevertheless, they suffer from significant performance penalties imposed by intensive KV cache accesses due to limited PCIe bandwidth. 
To address these issues, we propose \emph{InstInfer}, a novel LLM inference system that offloads the most performance-critical computation (i.e., attention in decoding phase) and data (i.e., KV cache) parts to Computational Storage Drives (CSDs), which minimize the enormous KV transfer overheads. 
InstInfer designs a dedicated flash-aware in-storage attention engine with KV cache management mechanisms to exploit the high internal bandwidths of CSDs instead of being limited by the PCIe bandwidth. The optimized P2P transmission between GPU and CSDs further reduces data migration overheads.
Experimental results demonstrate that for a 13B model using an NVIDIA A6000 GPU, InstInfer improves throughput for long-sequence inference by up to 11.1$\times$, compared to existing SSD-based solutions such as FlexGen.
\end{abstract}

\date{\today}
\section{Introduction}
\label{sec:intro}
%The prevalence of Large Language Models (LLMs) has ushered in a new era of generative AI \cite{}%. Models such as GPT-4 \cite{}, Bard \cite{}, and LLaMA \cite{} , which have become indispensable productivity tools,
Large language models (LLMs) and their underlying transformer architecture have revolutionized AI and have become the bedrock of many emerging applications, widely used in domains such as chatbot~\cite{achiam2023gpt4}, summarization~\cite{Wang2023ElementawareSW}, and code generation~\cite{Nijkamp2022CodeGenAO}.
Most of these LLMs are built based on the transformer architecture~\cite{vaswani2017attention} with enormous numbers of parameters and perform inference in an autoregressive manner consisting of two phases: \emph{prefilling} phase and \emph{decoding} phase. 

\begin{figure}
    \centering
    \includegraphics[width=0.99\linewidth]{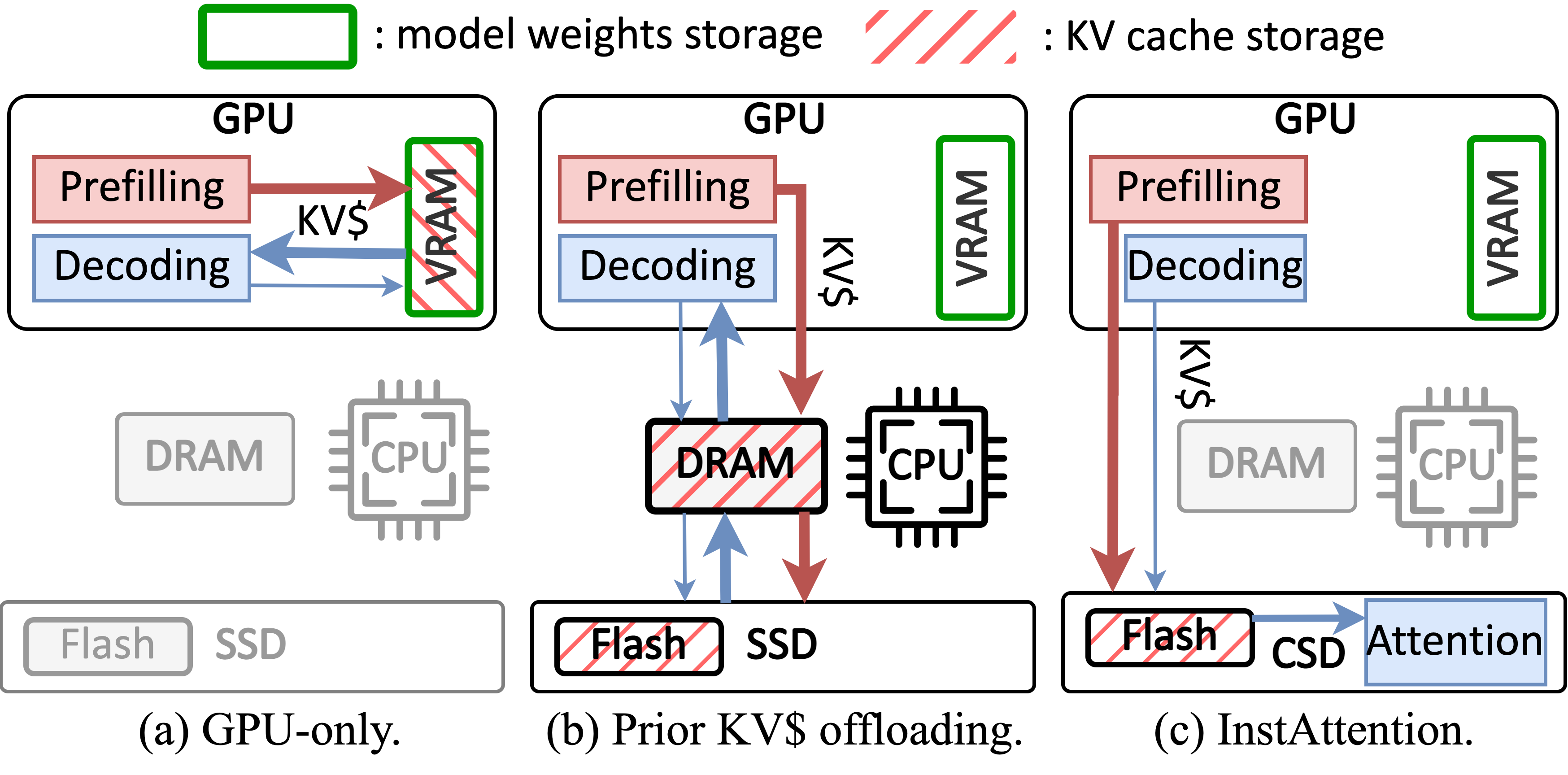}
    \vspace{-17pt}
    \caption{Comparison of different LLM inference architectures.}
    \label{fig:intro}
    \vspace{-15pt}
\end{figure}

% Previous research indicates that the prefilling phase is compute-bound, dominated by General Matrix Multiply (GeMM) operations, while the decoding phase is memory-bound, primarily involving General Matrix-Vector Multiply (GeMV) operations, with cached key-value tensors (KV cache) to reduce computational complexity \cite{}.
Previous research \cite{wu2024pim, heo2024neupims, zhong2024distserve, agrawal2024sarathi, zhao2024llmpq} indicate that the prefilling phase is compute-bound, while the decoding phase turns memory-bound owing to its key technique, called \emph{KV cache}. The KV cache is the cached key-value tensors, which store intermediate results from previous tokens. It significantly reduces computational complexity by allowing the model to reference past information efficiently without time-consuming recomputations \cite{kwon2023efficient, gao2024cachedattention}.
Considering the computing and bandwidth requirements, leveraging the extensive computing power and large VRAM bandwidth to accelerate LLM inference is the mainstream choice. As illustrated in Figure \ref{fig:intro}(a), the \texttt{GPU-only} architecture stores all weights and KV caches in the GPU VRAM, meanwhile leveraging the GPU to accelerate both prefilling and decoding phases. 

% Considering the significant differences between the two phases, there exist multiple architectures for LLM inference. The conventional approach solely deploys both prefilling and decoding phases (\textbf{P-D} for short) on homogeneous GPU clusters \cite{}, as shown in Figure \ref{}. 
% However, as the context length in LLM inference increases \cite{} and the batch size of a single inference grows \cite{}, the memory footprint of the KV cache escalates \cite{}. 
% Therefore, as the decoding phase predominately relies on the large volume of KV cache access, both the memory capacity and the bandwidth become critical challenges to the homogeneous P-D architecture. 
% Several works propose the coarse-grained disaggregated P-D architecture \cite{patel2023splitwise, zhong2024distserve}, which assigns the P-phase to GPUs with more computing resources and the D-phase to cost-efficient GPUs in terms of memory bandwidth. This approach only meets the bandwidth demands for the D-phase, but not the capacity issue. 
% In fact, the cost of deploying more GPUs to store the KV cache can become prohibitively high, even surpassing the cost of storing the model weights themselves. For instance, for a mid-sized LLM with 13 billion parameters, a batch size of 32, and a sequence length of 4K tokens, approximately 100GB of KV caches are required, which is 4.2$\times$ the size of the model itself. 
% Since the decoding phase is memory-bound and heavily reliant on KV cache I/O, storing the KV cache in GPU memory significantly increases both the storage cost and the idle rate of GPU computing resources.

Current LLM inference services can be categorized into online and offline scenarios. Online inference prioritizes low latency and typically accept shorter sequences from users \cite{fu2024serverlessllm}; whereas offline reasoning usually deals with longer sentences, accepting longer delays in exchange for higher throughput \cite{sheng2023flexgen}.
As LLMs continue to evolve, which have pushed the boundaries toward longer context reasoning \cite{lin2024infinite, jin2024longlm} and larger inference batches \cite{sheng2023flexgen, juravsky2024hydragen}, the memory footprint of their associated KV cache in offline-inference escalates drastically \cite{Wonbem2024InfiniGen}, introducing substantial challenges in storing them efficiently.
%Specifically, due to the extensive computational requirements, deploying GPUs to accelerate LLM inference is the mainstream choice, as illustrated in Figure \ref{fig:intro1} \cite{}. 
The situation gets more severe in resource-constraint scenarios such as edge computing or personal devices \cite{li2024personal, yin2024llmsystem, alizadeh2023llminaflash}. To be specific, the financial burden of deploying additional GPUs to accommodate the expansive KV cache can become exorbitantly high, potentially exceeding even the costs associated with storing the model weights.
%the cost of deploying more GPUs to store the KV cache can become prohibitively high, even surpassing the cost of storing the model weights themselves. 
For instance, a mid-sized LLM with 13 billion parameters, operating at a batch size of 32 and 4K tokens, necessitates approximately 100GB of KV cache. 
This volume is 4.2$\times$ the size of the model itself.%, posing significant challenges for cost-effective deployment in environments where resources are limited. %The typical \texttt{GPU-only} architecture exacerbates this issue, as it solely scales up by adding more GPUs to extend VRAM capacity for additional KV cache. 
% This approach not only significantly increases storage costs but also leads to severe wastage of extra computing resources.
% For instance, for a mid-sized LLM with 13 billion parameters, a batch size of 32, and a sequence length of 4K tokens, approximately 100GB of KV cache is required, which is 4.2$\times$ the size of the model itself. 
% The \texttt{GPU-only} architecture solely scales up to more GPUs to extend the VRAM capacity for more KV cache, which not only significantly increases the storage cost but also leads to severe wastage of extra computing resources.
%Since the decoding phase is memory-bound and heavily reliant on KV cache I/O, storing the KV cache in GPU memory significantly increases both the storage cost and the idle rate of GPU computing resources.

% \begin{figure}
% \vspace{-8pt}
% \centering
% %\def\subfigcapskip{0pt}
% \subfloat[GPU-only.]{\label{fig:intro1}\rotatebox{0}{\includegraphics[width=0.33\linewidth]{figs/intro_1.pdf}}}
% \subfloat[Prior KV\$ offloading.]{\label{fig:intro2}\rotatebox{0}{\includegraphics[width=0.33\linewidth]{figs/intro_2.pdf}}}
% \subfloat[InstInfer.]{\label{fig:intro3}\rotatebox{0}{\includegraphics[width=0.33\linewidth]{figs/intro_3.pdf}}}
% \vspace{-5pt}
% \caption{Comparison of different LLM inference architectures.}
% \vspace{-15pt}
% \end{figure}

To mitigate the storage costs associated with the KV cache, several approaches (e.g., DeepSpeed-MII \cite{holmes2024deepspeed} and FlexGen \cite{sheng2023flexgen}) have adopted more economical solutions, which offload the KV cache to host memory or cheaper SSDs for throughput-oriented offline inference, as shown in Figure \ref{fig:intro}(b). 
Before the GPU begins the decoding phase of inference, the KV cache is first loaded from SSDs to the memory and then to the GPU via PCIe buses.
%When KV cache volumes are relatively small, they transfer the KV cache from the memory to the GPU via PCIe buses before the GPU begin the decoding phase of inference \cite{}. For longer contexts or larger batches, the KV cache is firstly loaded from SSDs to the memory and then to the GPU \cite{}.
However, this offloading strategy introduces severe performance penalties. In particular, the PCIe bandwidth between host memory and GPUs is substantially lower than the bandwidth within GPU VRAM \cite{Wonbem2024InfiniGen}, while the bandwidth of SSDs is even lower. Additionally, the lack of direct datapath between the SSD and GPU and the complicated host-oriented storage software stack further exaggerate the performance penalty of SSD-offloading solutions.
Unlike the compute-bound prefilling phase, the memory-bound decoding phase critically depends on KV cache I/O, as it requires frequent transfers of large KV cache volumes between the storage media and GPUs. 
This dependence makes data movement over a narrow PCIe bus a new performance bottleneck.

To address the storage cost and bandwidth issues associated with KV cache, Computational Storage Drives (CSDs) \cite{mansouri2022genstore, yang2023lambda, lee2022smartsage} become a promising and cost-effective solution. Built on modern high-capacity SSDs, CSDs integrate computational resources such as FPGA accelerators internally. They present two advantages: 1) The storage cost of CSDs is comparable to that of SSDs \cite{koo2017summarizer, cao2020polardb}. Unlike the expensive GPU and host memory, the affordable storage capacity of SSDs can satisfy substantial capacity requirements of KV cache for long-context and large-batch scenarios. 2) Modern SSDs typically aggregate the throughput of all flash chips to deliver high internal bandwidth (tens of GB/s) \cite{wang2024beacongnn, pan2024flagger, zhang2020zng}, which is significantly higher than the external PCIe bandwidth of SSDs (3$\sim$6 GB/s) \cite{myung2020efficient, haas2023modern}. 
% By aggregating multiple CSDs, a high bandwidth can be achieved at a low cost. 
Offloading inference tasks to the computing engines within CSDs allows operands to leverage the high internal bandwidth directly. This bypasses the bandwidth-limited external PCIe bus, thereby meeting the KV cache bandwidth requirements.

Nevertheless, due to power consumption and cost constraints \cite{hadian2018towardscsd}, the computational power of CSDs is 2$\sim$3 orders of magnitude weaker than GPUs, making it ineffective to accelerate the entire inference tasks (cf. Section \ref{sub:motiv_oppo}). Instead, CSDs must collaborate with GPUs to accelerate LLM inference as a novel heterogeneous system, which, while seemingly straightforward, presents significant challenges:

\noindent $\bullet$ \emph{Coarse task partitioning between the GPU and CSD.} Existing heterogeneous LLM inference solutions typically disaggregate the prefilling and decoding phases \cite{patel2023splitwise, zhong2024distserve}. However, considering the much lower computing power of the CSD compared to the GPU, the entire decoding task exceeds the computing capability of the CSD, which becomes a new performance bottleneck.

\noindent $\bullet$ \emph{Significant bandwidth gap between CSD and GPU.} Both the external PCIe bandwidth and the internal bandwidth of CSD are still much lower than the GPU. For memory-bound decoding phase inference, reducing the data migration overheads remains necessary.

\noindent $\bullet$ \emph{Discrepancy between flash and memory access patterns.} %Flash can only be accessed in page granularity, with high access latency (30$\sim$90us) \cite{} and involving 
NAND flash accesses necessitate page granularity, high access latency, and complex multi-layer address translation mechanisms including the host file system and the Flash Translation Layer (FTL) \cite{hsieh2013multiftl}. Therefore, existing KV cache management mechanisms designed for memory (e.g., vLLM \cite{kwon2023efficient}) cannot be directly applied within the CSD.

Tackling the aforementioned challenges, we propose \emph{InstInfer}
\footnote{InstInfer is planned to be fully open-source.},
a novel LLM inference system based on in-storage computing and flash-based KV cache offloading, which effectively addresses both the storage cost and bandwidth limitations in traditional offline-inference schemes incurred by enormous KV cache volume, as illustrated in Figure \ref{fig:intro}(c). 
Specifically, to alleviate the computing burden of CSDs, InstInfer only offloads the most performance-critical \emph{decoding-phase attention} computations during long-context inference to CSDs, while leveraging the GPU to execute the remaining inference tasks.
To mitigate the computation power and bandwidth gap between CSD and GPU, InstInfer designs dedicated flash-aware in-storage computation engines with algorithm-hardware co-design for attention operators, which effectively lower the computation intensity and KV cache demands.
%Specifically, to minimize the KV cache transfer overheads while alleviating the computing burden of CSDs, InstInfer accommodates KV cache in flash chips to reduce storage costs. Only the most performance-critical \emph{decoding-phase attention} computations are offloaded during long-context inference to CSDs. %The remaining inference tasks continue to be performed by the GPU. 
InstInfer further proposes a KV cache-oriented FTL design to enable efficient KV cache access on the flash chips. 
% Through the dedicated flash-aware in-storage computation engine and algorithm-hardware co-design for attention operators, InstInfer effectively mitigates the computation power and bandwidth gap between CSD and GPU. 
The GPU and CSDs are directly connected via PCIe peer-to-peer DMA \cite{GPUDirectRDMA2017}, bypassing the host to avoid extra copies. Meanwhile, the works as the control plane, which only manages user requests, task scheduling, and data movement coordination. 
%LLM inference bottleneck caused by large KV cache IO in traditional offloading systems at minimal cost. 
%Therefore, through GPU-CSD heterogeneous collaborative inference, InstInfer achieves high-throughput LLM inference in large KV cache scenarios at minimal cost, effectively resolving the trade-off challenge between KV cache storage cost and bandwidth in traditional LLM inference systems.
To the best of our knowledge, InstInfer is the \emph{first work} to exploit \emph{CSDs} to address the performance penalty incurred from KV cache offloading. Experimental results show that for a 13B model with an NVIDIA A6000 GPU \cite{NVIDIA6000}, the throughput for long-sequence inference is improved by up to 11.1$\times$, compared to FlexGen.

\label{sub:llmbasic}
\begin{figure*}
    \centering
    \includegraphics[width=0.99\linewidth]{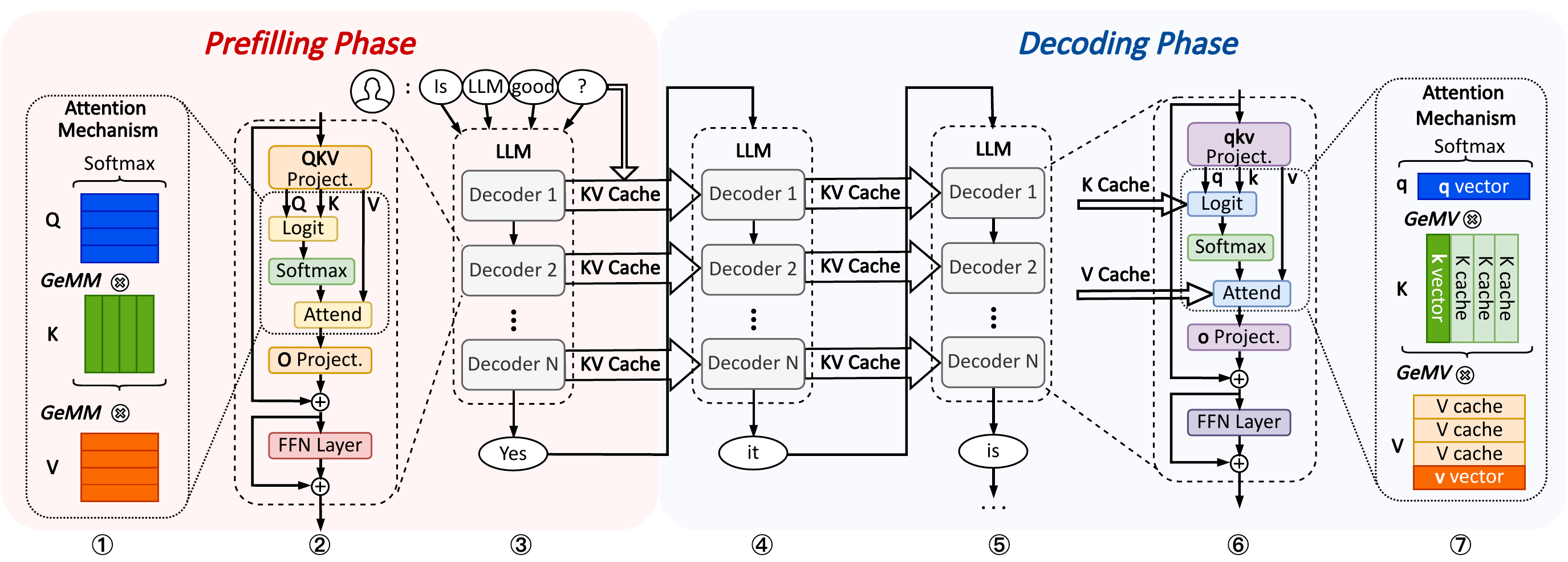}
    % \vspace{-40pt}
    \caption{General architecture of LLM and the inference flow.}
    \label{fig:LLMinfer}
    \vspace{-10pt}
\end{figure*}

The main \textbf{contributions} of this work are as follows:

\noindent $\bullet$ \emph{Pioneering CSD-based or GPU-CSD heterogeneous LLM inference system for long contexts:} Our detailed analysis reveals that the decoding-phase attention is the most critical performance bottleneck, due to the restricted PCIe bandwidth to access large KV caches. It exhibits extremely low arithmetic intensity, rendering GPU acceleration ineffective.
To address this, InstInfer offloads both KV cache and memory-intensive decoding-phase attention to the CSD, which exploits the high aggregated bandwidth of flash chips. Consequently, the data migration overheads are effectively mitigated by up to 94.0\%, while the prefilling phase overheads are further alleviated by the optimized peer-to-peer DMA mechanism.

\noindent $\bullet$ \emph{Hardware-algorithm co-designed in-storage attention engine:} To effectively bridge the bandwidth and computation power gap between GPU and CSD, we propose the bandwidth-efficient \emph{SparF} algorithm, which not only reduces the computing intensity but also minimizes the required KV cache volume during the decoding-phase attention while maintaining accuracy. Considering the page granularity of flash accesses, InstInfer incorporates a \emph{dual-step loading} strategy to manage the sparsity in the sequence: initially at the page level and subsequently at the token level. We further implement the in-storage SparF attention engine in hardware kernels, with fine-grained parallelism design to conceal the long access latency of flash chips, thereby improving the inference efficiency.

\noindent $\bullet$ \emph{KV cache-oriented FTL design for efficient retrieval:} With the SparF algorithm identifying sparsity patterns in both tokens and hidden embeddings, the resultant random access to KV caches in the flash chips presents a significant challenge. InstInfer confronts this issue by introducing \emph{dual address mapping} mechanisms tailored for token-indexed and hidden embedding-indexed KV caches, respectively. We further meticulously organize KV cache tensors into groups that align with flash page sizes and distribute them across multiple flash blocks and chips in a stridden fashion for each attention head, thus exploiting the inherent high parallelism.

\section{Background}
\label{sec:background}
\subsection{LLM Inference Basics}
\noindent \textbf{LLM Architecture.} Mainstream Large Language Models (LLMs) predominantly utilize a decoder-only transformer architecture \cite{Yang2024Harnessing, wang2024visionllm, zhang2022opt, touvron2023llama}. This architecture primarily comprises multiple stacked decoder blocks, each consisting of a self-attention module and a Feed-Forward Neural Network (FFN) module. Blocks \ding{173} and \ding{174} in Figure \ref{fig:LLMinfer} illustrate this process. For a given sequence of inputs $X=[x_1, ..., x_s]$, each decoder block applies linear transformations to $X$ with the parameter matrices, mapping $X$ into three embedding matrices: Q, K, and V, through GeMM computations.
% \begin{equation}
%     \label{QKVgen}
%     Q = W_Q X, \; K = W_K X, \; V = W_V X
% \end{equation}
Subsequently, the attention mechanism \cite{vaswani2017attention} is performed to capture the semantic context of the sentence: $Attention(Q, K, V) = softmax(\frac{QK^T}{\sqrt{d_k}})V$.
% \begin{equation}
%     \label{Attention}
%     Attention(Q, K, V) = softmax(\frac{QK^T}{\sqrt{d_k}})V
% \end{equation}
To enhance the ability of the vanilla attention mechanism to capture various aspects of the context, the Multi-Head Attention (MHA) \cite{vaswani2017attention} further divides the QKV matrices into multiple smaller matrices. This approach allows the model to focus on different parts of the input sequence simultaneously.
% Some variants of the vanilla attention mechanism divide the QKV matrices into multiple smaller matrices, a technique known as Multi-Head Attention (MHA) \cite{}. This approach allows the model to focus on different parts of the input sequence simultaneously, enhancing its ability to capture various aspects of the context.
The resultant attention output is then subjected to a linear transformation via another O matrix before being processed by the FFN layer. This output is then fed into the next decoder block as input. After the input passes through all the decoder blocks, the final predicted token is generated.

\noindent \textbf{Auto-regressive Inference.} LLM inference leverages an auto-regressive approach \cite{Wonbem2024InfiniGen}, which is divided into the prefilling phase and the decoding phase (cf. blocks \ding{174}$\sim$\ding{176} in Figure \ref{fig:LLMinfer}). During the prefilling phase, the LLM processes all the tokens of the input prompts in parallel to generate the first predicted output token $x_{s+1}$. This token is then appended to the existing input prompt sequence to generate the new input sequence $[x_1, ..., x_s, x_{s+1}]$. In the decoding phase, the LLM predicts one new output token at a time based on this sequence, and gets a new predicted token $x_{s+2}$. This process repeats iteratively until an End-of-Sequence (EOS) token is generated or the model reaches its context limit.

\subsection{KV Cache}
\label{sub:kvcache}
\noindent \textbf{Recomputation reduction.} During the decoding phase of LLM inference, the input for each inference step consists of the entire sequence generated so far. Consequently, the attention operation requires repeated calculations of the QKV matrices for all the previous tokens, resulting in a computational complexity of $O(s^2)$ per iteration \cite{vaswani2017attention}.

A highly effective method to alleviate the computational bottleneck in LLM decoding is the KV cache \cite{kwon2023efficient} (cf. blocks \ding{177} and \ding{178} in Figure \ref{fig:LLMinfer}). By caching the KV matrices for the already generated tokens in the GPU memory, redundant calculations can be avoided. Thus, when computing the new attention output, only the KV vectors for the new token need to be calculated. This optimization reduces the attention calculation in each decoding step from GeMM to GeMV, thereby lowering the computational complexity of the attention layer from $O(s^2)$ to $O(s)$. However, as the context length of LLMs increases, storing the KV cache consumes substantial memory space and imposes high I/O demands during the decoding phase \cite{Wonbem2024InfiniGen}.

\noindent \textbf{Sparse Attention.} To further reduce the memory access demands of attention, sparse attention has become a commonly adopted method \cite{chen2021scatterbrain, chaudhari2021attentive}. This approach is based on the observation that within a text sequence, the importance of different tokens varies; in a fully connected attention mechanism, some weak connections contribute minimally to the final attention output and can be disregarded. By reducing the number of KV vectors for tokens to be calculated and stored, sparse attention opens up possibilities for decreasing the computational and storage overhead of the KV cache.

Prior works have proposed various sparse attention algorithms \cite{liu2024scissorhands, zhang2024h2o, ribar2023sparq, liu2023dejavu, tang2024razorattention, zhang2024q}, among which SparQ Attention \cite{ribar2023sparq} is optimized for bandwidth-efficient inference scenarios. Unlike other algorithms that compute the complete attention score, it approximates the attention score based on the $r$ largest hidden embedding values in the query (Q) vector. It then identifies the top-$k$ most important tokens based on the approximated attention score with full hidden embeddings to calculate the attention output. To compensate for the omitted value (V) tensors, the V tensors are weighted and averaged, merging them into the final attention output. On multiple datasets, SparQ Attention reduces the bandwidth requirement for KV cache transmission during the decoding phase by up to 7/8 while maintaining accuracy. However, the SparQ attention algorithm only reduces the bandwidth demand for KV cache access but requires 1.5$\times$ larger KV cache memory footprints. This is because it needs to index the key (K) cache by both token dimension and hidden embedding dimensions, limiting its applicability in memory-constrained scenarios.

\begin{figure}
    \centering
    \includegraphics[width=0.95\linewidth]{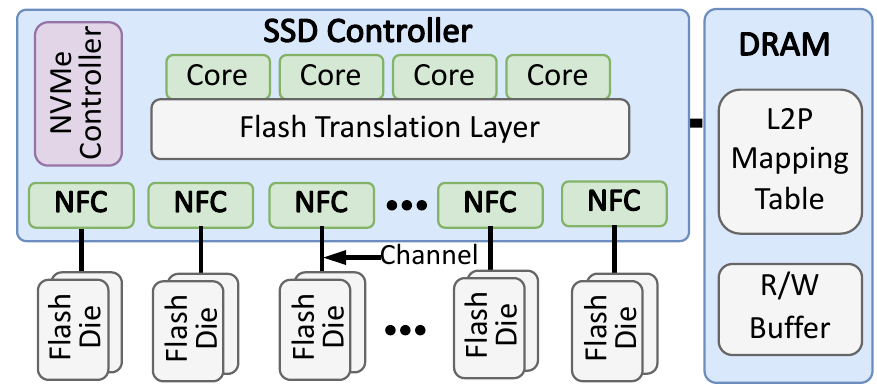}
    % \vspace{-40pt}
    \caption{A typical SSD architecture.}
    \label{fig:ssd}
    \vspace{-15pt}
\end{figure}

\subsection{SSD and In-Storage Computation}
\label{sub:ssd}
\noindent \textbf{SSD Basics.} Figure \ref{fig:ssd} illustrates the internal organization of a modern NAND-flash-based solid-state drive (SSD) \cite{mao2017improving}, which comprises three main components: NAND flash dies, an SSD controller, and a DRAM module. One or more dies share command/data buses, known as \emph{channels}, to connect to the SSD controller. %Each channel can transfer data independently of the others. 
Each die is subdivided into 2$\sim$4 planes, and each plane contains thousands of blocks. A block is further divided into hundreds of pages, typically ranging from 4KB to 16KB in size \cite{desnoyers2012analytic}. Pages are the smallest read/write units of flash chips. Before data can be written to flash pages, the flash memory needs to be erased at the block level \cite{agrawal2008design}.

The SSD controller generally consists of three parts: a general-purpose processor running the flash translation layer (FTL), an NVMe controller, and NAND flash controllers (NFCs). The FTL is responsible for managing the logical-to-physical address mapping of the data stored in the flash dies and scheduling tasks on the NAND flash. The NVMe controller facilitates communication with the host via the NVMe protocol \cite{NVme2.0d}, while the NFCs manage communication with the flash backend. Each NFC operates on a flash channel for independent data transfers. Modern SSDs typically feature 8$\sim$16 flash channels, with each channel capable of transferring data at rates of 1$\sim$2GB/s \cite{zhang2020zng}. Consequently, the aggregated bandwidth of flash channels can reach tens of GB/s, significantly exceeding the external PCIe bandwidth of SSDs (3$\sim$6GB/s) \cite{980pro}.
The DRAM within the SSD functions as a temporary buffer for data being read from or written to the flash dies. It also maintains the logical-to-physical (L2P) mapping table and other metadata for the FTL.

\noindent \textbf{Computational Storage Drive.} To leverage the high internal bandwidth of modern SSDs, the computational storage drive (CSD) employs in-storage computation techniques by integrating computing engines, such as ARM cores, NPUs, or FPGA chips, within the SSD \cite{koo2017summarizer, liang2019cognitive, mansouri2022genstore}. This integration endows the SSD with computing capabilities, enabling it to perform data processing tasks directly within the storage device. It is worth noting that, to fully utilize the high flash channel bandwidth, %the computing engine must be placed near the flash dies or NFCs, 
it would be better to place the computing engine near the flash dies or NFCs rather than being connected to the SSD through a PCIe switch (i.e., Samsung SmartSSD \cite{soltaniyeh2022smartssd}). This in-storage computing architecture is employed in InstInfer to harness the substantial bandwidth necessary for LLM inference with a large KV cache.

\section{Challenges and Opportunities}
\label{sec:motivation}
\subsection{Limitations of Conventional KV Cache Offloading}
\label{sub:motiv_limit}
\noindent \textbf{KV Cache Analysis.}
Nowadays, the context length of the LLM inference serving system is continuously increasing\cite{jin2024longlm, lin2024infinite}.% The commonly adopted ShareGPT dataset \cite{shareGPT} includes long conversation contexts exceeding 2K tokens accounting for X\% of the data. Furthermore, according to \cite{}, current LLM serving systems are supporting contexts up to 128K, which leads to an enormous KV cache size. 
Furthermore, in order to enhance the GPU utilization rates, LLM inference systems typically batch many requests in a single iteration to leverage the parallel computing capabilities of GPU \cite{chen2021re, zhang2019flashgpu, pan2023bcbench}, which also extends the KV cache size.
%Nevertheless, increasing the inference batch size also extends the KV cache size.
Assuming that $b, s, p$ denote the batch size, sequence length, and model parameter size, respectively, the KV cache size stored in the FP16 format is $4bsp$, while the model weight size in FP16 format is only $2p$. For a 2K-length sequence with batch size 128, the OPT-13B model occupies about 24GB for its model weights and generates up to 200GB KV caches. For larger models like OPT-175B, the model weights occupy 325GB, while the KV cache reaches up to 2.63TB. Given that the precious GPU memory will be primarily allocated for storing the model weights and activations, the KV cache tends to be offloaded to host memory or SSDs for cost-effectiveness, depending on the sizes.

\begin{figure}
    \vspace{5pt}
    \centering
    \includegraphics[width=0.99\linewidth]{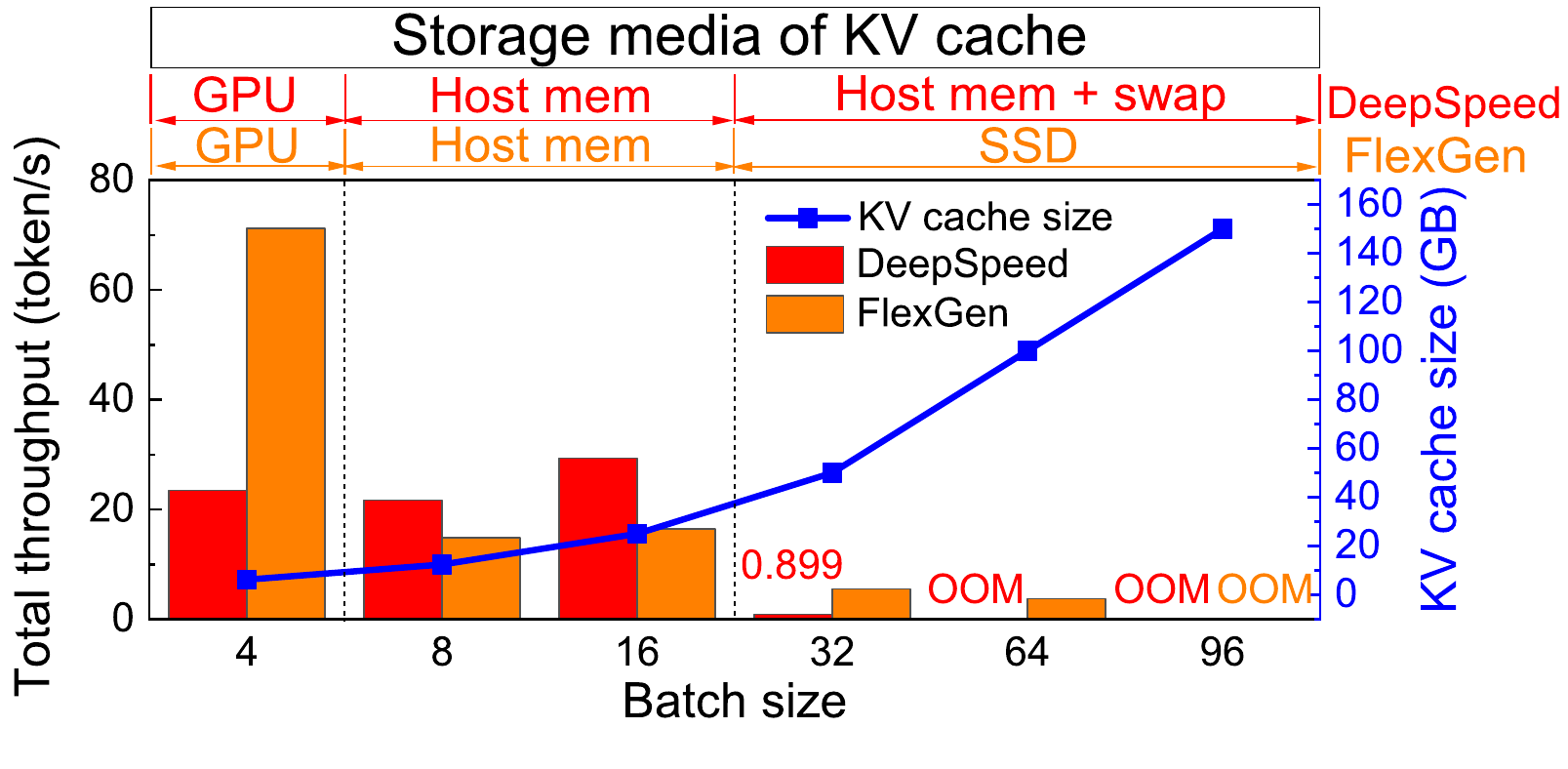}
    \vspace{-15pt}
    \caption{Throughput of DeepSpeed and FlexGen.}
    \label{data:motiv_thrpt}
    \vspace{-10pt}
\end{figure}

\noindent \textbf{Performance Degradation With Offloading.}
As the PCIe bandwidth between host memory or SSDs and the GPU is significantly lower than that of GPU memory, prior offloading schemes lead to a noticeable decline in inference performance. To demonstrate this, we evaluate the inference throughput of two latest KV cache offloading systems, DeepSpeed \cite{holmes2024deepspeed} and FlexGen \cite{sheng2023flexgen}, in a long-context scenario of the OPT-13B model. We evaluate them with different batch sizes on an NVIDIA A6000 GPU, which possesses 48GB GPU memory. Both the input and output sequence lengths are set to 1024 tokens. As depicted in Figure \ref{data:motiv_thrpt}, both Deepspeed and FlexGen exhibit performance drop as batch sizes increase: Deepspeed at batch sizes 8 and 32, and FlexGen at batch sizes 8 and 64. These drops occur because the KV cache size exceeds the available GPU memory, necessitating offloading first to host memory and subsequently to SSD. Note that Deepspeed does not support SSD offloading; consequently, at a batch size of 32, kernel swapping from host memory to SSD occurs, leading to a 97.01\% performance decline.
While increasing the batch size within the same memory tier enhances throughput, offloading the KV cache to secondary storage significantly degrades the performance.

\begin{figure}
    \centering
    \includegraphics[width=0.99\linewidth]{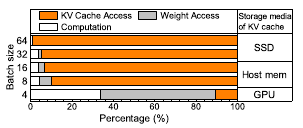}
    \vspace{-15pt}
    \caption{Latency breakdown of FlexGen decoding.}
    \label{data:motiv_brkd}
    \vspace{-10pt}
\end{figure}

To further elucidate the source of the performance penalty from offloading KV caches, we analyze the decoding-phase latency of FlexGen across different batch sizes, as illustrated in Figure \ref{data:motiv_brkd}.
For smaller batch sizes (4, 8), where all the KV caches fit within the GPU, the primary bottleneck is \texttt{Weight Access}. However, as the batch size increases and the KV caches are offloaded to memory or SSD, the overhead from \texttt{KV Cache Access} escalates to as high as 98.94\%. This substantial increase underscores the need for new solutions to address the significant performance challenges introduced by KV cache offloading.

\subsection{Offloading Opportunities with CSD}
\label{sub:motiv_oppo}
We discovered that compared to memory and NVMe SSDs, offloading KV caches to the flash chips within the CSD can directly leverage the higher flash channel bandwidth to meet the demands of the LLM decoding phase using the internal computational units. However, the simple prefilling-decoding separation architecture proposed in prior works \cite{patel2023splitwise, zhong2024distserve}, which typically targets GPU-CPU separation or distribution across different GPUs, is not suitable for CSD offloading. This is primarily due to the significant differences in the characteristics of various operators and the much lower performance of CSD compared to GPUs. Consequently, it is challenging for CSDs to handle the entire inference task independently.

\begin{figure}
    \centering
    \includegraphics[width=0.99\linewidth]{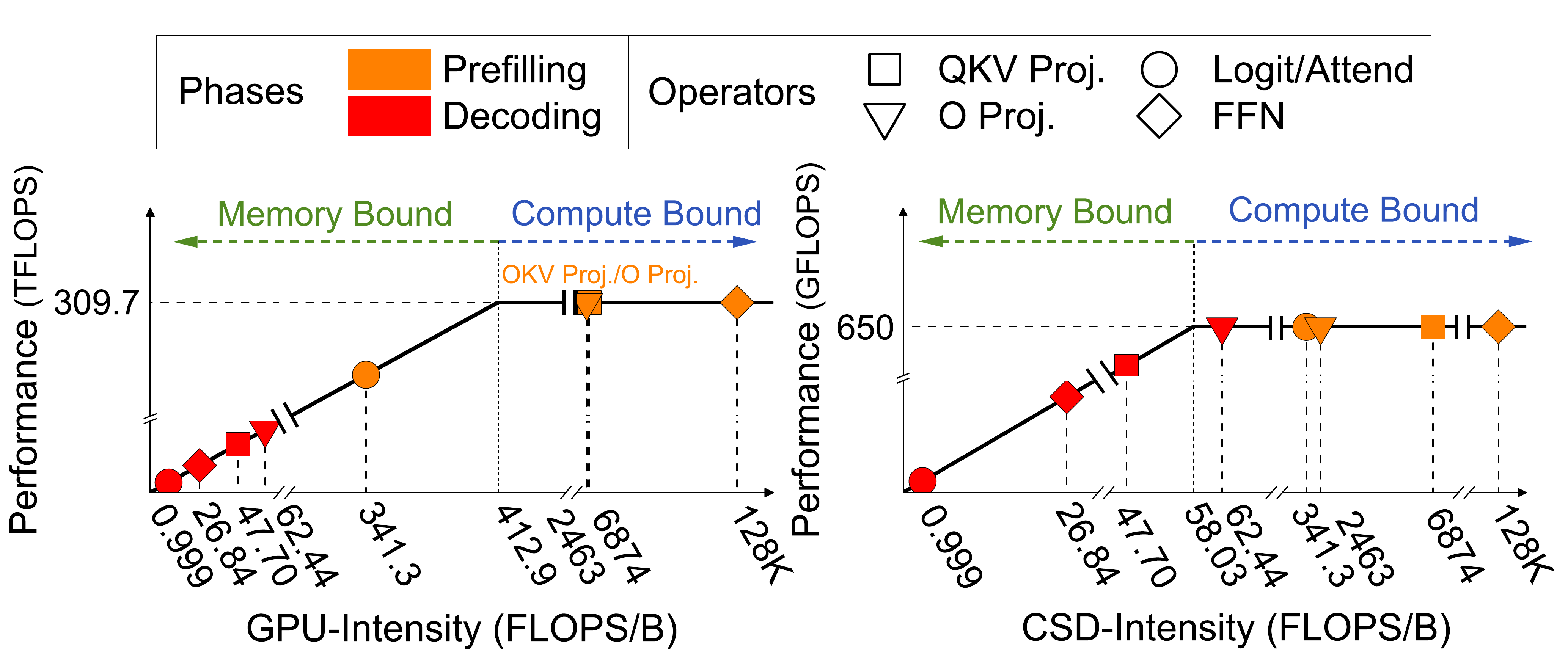}
    % \vspace{-40pt}
    \caption{Roofline models of different hardware.}
    \label{data:roofline}
    \vspace{-10pt}
\end{figure}

To minimize the computational load on the CSD and fully utilize its high internal bandwidth, a practical approach involves restructuring the scheme of task disaggregation. This can be achieved by offloading only memory-bound operators with low computing intensity and substantial KV cache I/O to the CSD, while retaining other operators on the GPU. 

To this end, we thoroughly analyzed the main operators involved in LLM inference, examining their patterns on both CSD and GPU. Figure \ref{data:roofline} illustrates the roofline models \cite{yuan2024roofline} of an NVIDIA A6000 GPU \cite{NVIDIA6000} and a Zynq7045 FPGA-based CSD \cite{zynq}. The hardware configurations are detailed in Section \ref{sec:impl}.
For the prefilling phase, the \texttt{QKV Proj.}, \texttt{O Proj.}, and \texttt{FFN} are extremely computing-intensive and should be placed on GPU. Although the attention operands (i.e., \texttt{Logit} and \texttt{Attend}) are memory-bound on the GPU, the limited computing power on CSD will severely constrain their performance. Therefore, the prefilling-phase attention should also remain on the GPU.
In contrast, the decoding-phase operators exhibit significantly different characteristics. For the decoding phase, although \texttt{QKV Proj.}, \texttt{O Proj.}, and \texttt{FFN} operands are memory-bound on the GPU and seem suitable for CSD-offloading, their operational intensities are near the maximum computing capability of CSD. This places a substantial burden on CSD's computing engine. Moreover, these operands rely solely on weight matrices for flat GeMM computations \cite{hong2024flashdecoding++}, independent of the KV cache on the flash chips. Conversely, the attention operands (\texttt{Logit} and \texttt{Attend}), which involve extremely low-intensity GeMV computations, require direct access to KV caches (see block \ding{178} in Figure \ref{fig:LLMinfer}). 
% If only the decoding-phase attention operands are offloaded to CSD, both the volume of KV cache transmitted over the limited external PCIe bus and the computational burden on the CSD will be minimized.
This motivates us to offload the decoding-phase attention operands to the CSD while retaining other processes on the GPU. This approach aims to significantly reduce KV cache transmission overheads and minimize the computational burden on the CSD.

\section{Design}
\label{sec:design}
\begin{figure}
    \centering
    \includegraphics[width=0.99\linewidth]{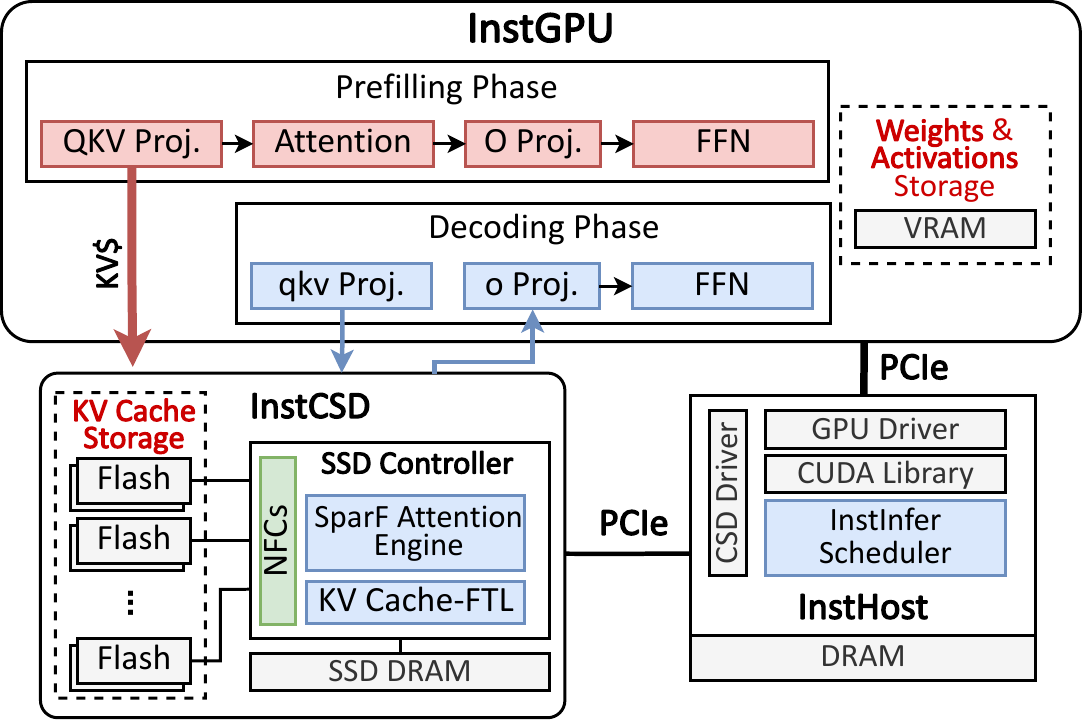}
    % \vspace{-40pt}
    \caption{Overview of InstInfer architecture.}
    \label{fig:InstInfer}
    \vspace{-20pt}
\end{figure}

% In this section, we first give an overview of InstInfer architecture (Section \ref{sub:overview}), and then present in detail how InstInfer computes the decoding-phase attention with the CSD (Section \ref{sub:compute}) and manages KV caches on the flash chips (Section \ref{sub:manage}). Finally, we present how the CSD is integrated with GPU as well as scaling up to the CSD arrays (or CSA) (Section \ref{sub:scale}).

\subsection{Overview of InstInfer}
\label{sub:overview}
Based on the insights presented in Section \ref{sec:motivation}, we propose \emph{InstInfer}, the first in-storage attention offloading system with general GPUs, tailored for offline LLM inference with long-context and large-batch.

The key idea of InstInfer lies in reducing both KV cache movement overheads and computational burden on the CSD, along with the corresponding flash-aware KV cache retrieval mechanism and co-design of attention operands. As illustrated in Figure \ref{fig:InstInfer}, InstInfer is primarily comprised of three hardware components: 1) \emph{InstCSD}, which executes decoding-phase attention computation and stores the large KV cache volumes; 2) \emph{InstGPU}, which performs other inference computations along with generating KV cache during the prefilling phase; and 3) \emph{InstHost}, which runs the software stack, scheduling inference tasks and orchestrating data transmission between the GPU and InstCSDs. 
% The general workflow with InstInfer is as follows.
% As user inputs arrive, the host CPU processes the requests, appending system prompts and tokenizing the inputs. Then the host batches the requests and sends the inputs to the GPU.
% The GPU initiates the inference process from the prefilling phase, and transmits the generated KV cache in a layer-wise way to InstCSD through peer-to-peer DMA. After InstCSD receives parts of the KV cache, it immediately begins to construct indexes and address mapping tables for the KV cache, and stores them in the backend flash chips. When the prefilling phase ends and the decoding phase begins, the first generated token will be transformed into $qkv$ vectors and then sent to InstCSD. Based on the SparF attention mechanism, InstCSD first identifies the primary embedding in the $q$ vector, and 

Given that for KV cache in the CSD, the storage requirement is significantly less than the demanding bandwidth, we propose the \emph{SparF Attention} mechanism, an enhanced version to the traditional SparQ algorithm \cite{ribar2023sparq} (cf. Section \ref{sub:kvcache}), specifically tailored for flash storage to trade storage capacity for reduced computation and data transmission on the CSD. Considering the page granularity of flash access, SparF Attention organizes tokens at a group level, which corresponds to the page size of flash chips to avoid wasting the flash channel bandwidth. The required KV cache tensors are thereby identified and fetched through a dual-step mechanism, initially in the coarse-grained group level and then in the fine-grained token level.
Based on the SparF Attention, we further design the hardware-based accelerator on InstCSD for computing the attention outputs at fine-grained parallelism, to effectively identify the sparsity pattern in the runtime. 
As SparF attention demands index KV cache in both hidden embedding and token, we propose two address-mapping mechanisms in the FTL of InstCSD for efficient retrieval of KV caches.
Through the InstCSD, only the $qkv$ vectors and attention output are transmitted between the GPU and InstCSD during the decoding phase. Furthermore, the KV caches generated by the GPU are transmitted to InstCSD through P2PDMA, bypassing the host memory and the burden from the filesystem. The KV cache transmission is executed in a layerwise way, overlapped with the inference computation to hide the transmission latency.
% , which significantly mitigates the bandwidth bottleneck of conventional KV cache offloading systems.  

% feels like my eyes are raped :(
\begin{algorithm}
    % \textsl{}\setstretch{1.8}
    \renewcommand{\algorithmicrequire}{\textbf{Input:}}
    \renewcommand{\algorithmicensure}{\textbf{Output:}}
    \caption{SparF Attention: Flash-aware Sparse q-Attention}
    \label{algo1}
    \begin{algorithmic}[1]
        \REQUIRE $\bm q, \bm{\bar v} \in \mathbb R^{d_h}, \bm {K,V} \in \mathbb R^{S\times d_h}, \bm{r,k,l, m,n} \in \mathbb N$
        \ENSURE $\bm{out}\in \mathbb R^{d_h}$
        \STATE $\bm i \gets [1\; {\rm \mathbf {if}} \; i \in {\rm argtopk} (|\bm{q} |, r) \; {\rm \mathbf {else}} \; 0]^S_{i=1}$
        % \STATE ${\rm load} \; \bm{K}^\top_{[:,\bm i]} \; \rm from \; flash$
        \STATE ${\rm load} \; \bm{K}^\top_{[\bm :,i_1]} \; {\rm \mathbf{if}} \;\bm {i}_{[\bm{m}\lfloor \frac{\bm i_1}{\bm m}\rfloor:\bm{m}\lceil \frac{\bm i_1}{\bm m}\rceil]}=\bm 0 $ 
        \STATE $\bm{K}^\top_{[\bm :,i]}\gets {\rm filter } \; \bm{K}^\top_{[\bm i_2,:]} \; {\rm \mathbf{if}} \; \bm i_2 = 0 $
        \STATE $\bm{\hat s} \gets {\rm softmax} \left(\bm{q_{[\bm i]}} \cdot \bm{K}^\top_{[:,\bm i]}/\sqrt{d_h\frac{||\bm q_{[\bm i]}||_1}{||\bm q||_1}}\right)$
        \STATE $\bm m \gets [{\rm 1 \; \mathbf{if}} \; i>S \; {\rm \mathbf{else} \; 0}]^S_{i=1}$
        \STATE $\bm j \gets [1\; {\rm \mathbf {if}} \; j \in {\rm argtopk} (\bm{\hat s}+\bm m, k) \; {\rm \mathbf {else}} \; 0]^S_{j=1}$
        \STATE $\alpha \gets {\rm sum}(\bm{\hat{s}}_{[\bm j]})$
        % \STATE ${\rm load} \; \bm{K}^\top_{[\bm j,:]} \; \rm from \; flash$
        \STATE ${\rm load} \; \bm{K}^\top_{[\bm j_1,:]},\bm V_{[\bm j_1,:]} \; {\rm \mathbf{if}} \;\bm {j}_{[\bm{n}\lfloor \frac{\bm j_1}{\bm n}\rfloor:\bm{n}\lceil \frac{\bm j_1}{\bm n}\rceil]}=\bm 0 $ 
        \STATE $\bm{K}^\top_{[\bm j,:]},\bm V_{[\bm j,:]}\gets{\rm filter } \; \bm{K}^\top_{[\bm j_2,:]},\bm V_{[\bm j_2,:]} \; {\rm \mathbf{if}} \; \bm j_2 = 0 $
        \STATE $\bm s\gets {\rm softmax}\left( \bm q \cdot \bm K^\top_{[\bm j,:]} / \sqrt{d_h} \right)$
        \STATE $\bm{out}\gets \alpha\bm s\cdot\bm V_{[\bm j,:]}+({\rm 1}-\alpha)\bm{\bar v}$
    \end{algorithmic}
    % \vspace{-30pt}
\end{algorithm}

\subsection{Compute Attention Outputs}
\label{sub:compute}
\noindent \textbf{Flash-Aware Sparse Attention.} Based on Section \ref{sub:motiv_oppo}, we observed that the decoding-phase attention operators (i.e., Logit and Attend) remain severely memory-bound on CSD due to their predominate reliance on the KV cache stored in the flash chips. 
Our solution leverages the inherent sparsity in the attention mechanism, which has been thoroughly exploited in prior works (cf. Section \ref{sub:kvcache}).
However, none of the prior spare algorithms consider the specific flash characteristics, rendering them unsuitable for CSD adoption. Specifically, compared with GPU VRAM or host DRAM, NAND-flash-based SSD has a much larger capacity with lower bandwidth, making the idea of trading storage capacity for bandwidth feasible. Furthermore, current sparsity algorithms generate an extensive number of random accesses to the KV caches, since different contexts present varying semantic relatedness among tokens. This leads to random accesses in flash chips based on the interrelatedness of tokens, resulting in significant write amplification \cite{hu2009write} and bandwidth wastage due to the page granularity of flash chip accesses.

To this end, we propose \emph{SparF Attention}, a flash-aware sparse q-attention algorithm, as illustrated in Algorithm \ref{algo1}. Building on the vanilla SparQ attention \cite{ribar2023sparq}, SparF Attention identifies the sparsity pattern between the current token (the $q$ vector) and existing sequence (the $K$ cache matrix) by selecting the top-$r$ hidden embedding values of the $q$ vector (step $1$ in Algorithm \ref{algo1}) to approximate the full attention score $\hat s$. During this process, the $K$ caches are loaded from flash chips indexed by the hidden embedding, corresponding to the sparsity found in the $q$ vector.
Based on the approximated attention score $\hat s$, SparF Attention further selects the top-$k$ largest tokens in $\hat s$ to approximate the final attention output, and load the full $K, V$ cache vectors of these tokens from flash. This time, both K and V caches are loaded from flash chips indexed by the tokens, corresponding to the sparsity pattern identified in step $6$ among the sequence. 
To fit with the page size of flash chips, the KV cache loading process is separated into two steps, as illustrated in steps $2-3,8-9$ in the Algorithm \ref{algo1}. The detailed data mapping scheme will be described in Section \ref{sub:manage}. In steps $2,8$, any flash page containing weak elements based on the results of $argtop-k$ in steps $1,6$ will not be fetched from the flash chips. After the coarse-grained sparse KV caches are fetched to the corresponding NFCs, the NFCs execute steps $3,9$ to further filter out the remaining weak units in the KV caches. This dual-step loading scheme thereby not only reduces the data volume transmitted through the flash channels but also alleviates the computation burden since all the weak units will be discarded.

\begin{figure}
    \centering
    \includegraphics[width=0.99\linewidth]{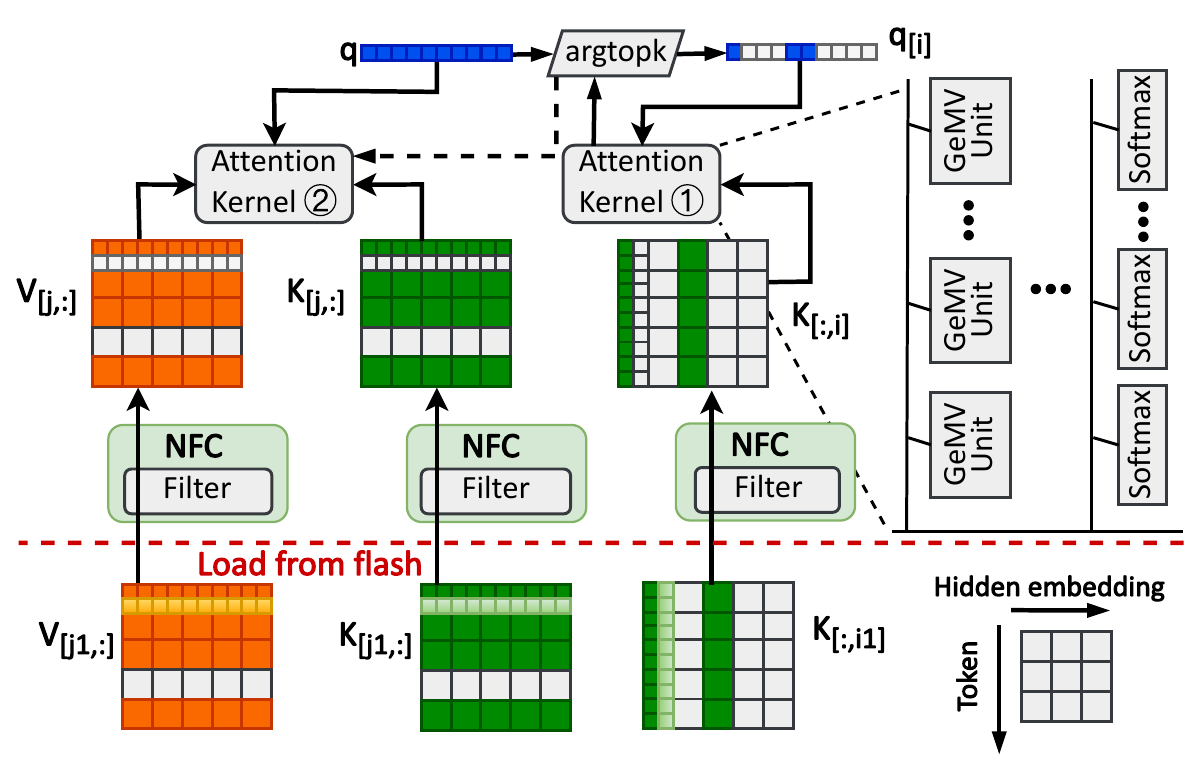}
    \vspace{-15pt}
    \caption{Workflow of SparF Attention engine on InstCSD.}
    \label{fig:instcsd}
    \vspace{-15pt}
\end{figure} 

\noindent \textbf{Hardware-Based Attention Engine.}
Based on the SparF Attention mechanism, we design the hardware-based attention engine on the InstCSD, which is integrated with the SSD controller. As depicted in Figure \ref{fig:instcsd}, the attention engine is primarily comprised of the attention kernels, argtopk unit, and the filters integrated within each NFC. Minor components, such as the summation unit, are omitted in the figure for simplicity. 
The $q$ vector is firstly sent into the \texttt{argtopk} unit to filter out the top-$r$ largest hidden embeddings. Subsequently, the top-$r$ indexes are sent to the NFC to fetch the $K_{[;,i]}$ caches, where each page might contain sparse units. These pages are then filtered when passing through the NFCs, which possess the fine-grained index information to filter out all the sparse units (cf. Section \ref{sub:manage} for details). $q_{[i]}$ and $K_{[;,i]}$ are then sent to the \texttt{Attention Kernel}\ding{172} to compute the approximate attention score, which is then sent to $argtopk$ unit again to filter the top-$k$ largest token indexes for the final attention output.
Based on the indexes, the $K_{[j;,]}$ and $V_{[j;,]}$ caches will be loaded from flash in page granularity, and filtered through the NFC as well. The sparse $q, K_{[j;,]}$ tensors are first loaded and sent to \texttt{Attention Kernel}\ding{173}, while the $V_{[j;,]}$ are loaded in parallel to hide the loading latency.
The two \texttt{Attention Kernel}s in the figure are identical, each composing multiple GeMV units and Softmax units to complete the attention computation involved in steps $4,10,11$ in Algorithm \ref{algo1}. During the execution, both attention kernels can be scheduled for the two attention computations in SparF Attention considering the real-time loads.

\begin{figure}
    \centering
    \includegraphics[width=0.99\linewidth]{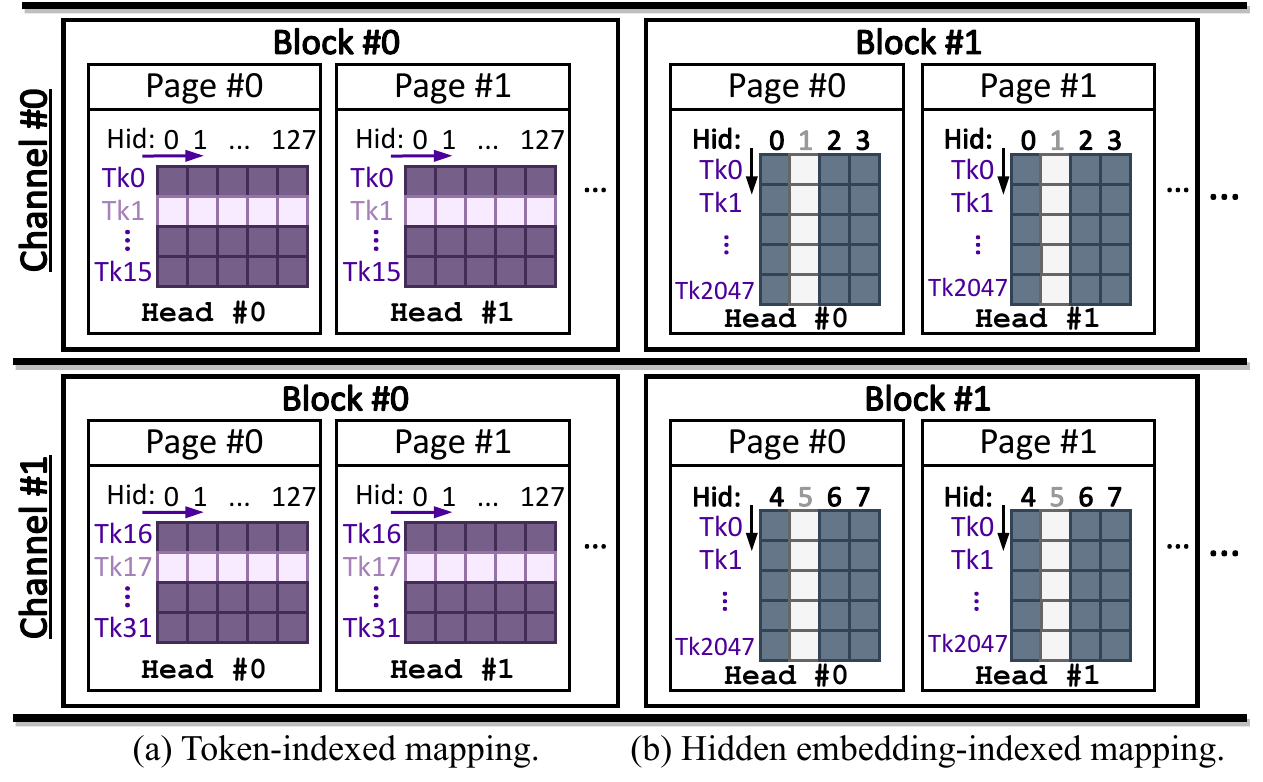}
    \vspace{-15pt}
    \caption{Schematic of different KV cache mapping schemes.}
    \label{fig:mapping}
    \vspace{-15pt}
\end{figure} 

\subsection{Manage and Transmit KV Caches}
\label{sub:manage}
To facilitate the SparF Attention mechanism within the CSD equipped with flash chips, it is necessary to enable token-index random access to the $V$ cache. Additionally, both token-index and hidden embedding-index random accesses are required for the $K$ cache. Therefore, considering the low storage cost of flash, we opted to store the $K$ matrices twice, each indexed in different orientations to optimize access efficiency. Additionally, we designed two sets of efficient address mappings with the dual-step loading mechanism. This approach enables random indexing and efficient flash memory access while significantly reducing write amplification.

\noindent \textbf{Token-Indexed Mapping.}
The token-indexed management is illustrated in Figure \ref{fig:mapping}(a). Specifically, we note that in mainstream large models such as OPT and LLaMA, each attention head has a hidden size of 128 and the storage format is FP16 \cite{zhang2022opt, touvron2023llama}, which implies a minimum reading granularity of 256B. Considering the 4KB page-granularity access of flash, randomly reading the KV cache can lead to performance degradation of up to 16$\times$ in conventional FTL.

To address this, we group $K$ or $V$ caches of 16 consecutive tokens into a \textit{group}, with each group stored within the same page and stridden across channels. For other configurations (e.g., larger head size or page size), the group size varies accordingly.
During the sparse pattern analysis process (e.g., step $5$ in Algorithm \ref{algo1}), a group will be ignored only if all its tokens rank below the top-$k$ threshold; otherwise, it is considered dense. This design ensures that KV cache reads are always at the page granularity, with data stridden across different channels to maximize the utilization of flash channel bandwidth.
Nevertheless, this approach inevitably fetches tokens that are considered sparse in the top-$k$ selection. 
To address this, we further propose a dual-step loading mechanism as shown in steps $2,3$ and $8,9$ of Algorithm \ref{algo1}, where the $\bf m,n$ represent the group size. The former loading step of complete token pages with sparse tokens is considered the first step.
%We integrate a filter in each NFC with stream processing capability.
After the group-wise sparse KV caches are fetched from flash chips to the buffer within the NFC, the filter further filters out the sparse tokens in the group, leaving only the strong units to be loaded to the on-chip buffer of the attention engine. 
% is it suitable to call them "non-zeros?"
Our tests imply that this group-based dual-step loading scheme generally maintains about half of the sparsity across various datasets during the first-step loading, while the second step reaches the full sparsity. Both $K$ and $V$ caches of the token index are stored and fetched in the same manner.

\noindent \textbf{Hidden Embedding-Indexed Mapping.}
For the hidden embedding-indexed access, the approach is relatively similar, as illustrated in Figure \ref{fig:mapping}(b).
Since each hidden embedding index requires continuous access to multiple tokens, we need to store the $K$ cache corresponding to multiple tokens consecutively within a single page, with each token occupying only one hidden embedding. For a 4KB page, each page can store 2K tokens, which is quite a large granularity for general LLM inference. We further adopt the two-step loading mechanism for the hidden embedding-indexed access, grouping 2-8 hidden embeddings into one flash page. 
%The order of different hidden embeddings strides across channels.
Therefore, the minimum storage granularity in this scenario is 256-1K tokens, which is feasible for both short conversations (less than 256 tokens) and long contexts (longer than 1K tokens). The group size can be dynamicallyadjusted based on the input length and largest context length of the model in the runtime.

\noindent \textbf{Batch Writing Requests.}
During the prefilling phase, after all the KV cache chunks of input tokens are transmitted and stored in the flash chips, the decoding phase still generates KV vectors for new tokens incrementally, which are continually transmitted to the CSD for storage. Owing to the page granularity of flash writes and the group granularity of tokens, the one-by-one generated tokens are required to first store the corresponding KV caches in the DRAM group buffer within the CSD. The KV caches of these tokens will be subsequently flushed back to the flash chips in the background when full.

To further mitigate write amplification during flash write operations, it is crucial to ensure that each write operation is at block granularity comprising several hundred pages \cite{hu2009write}. 
When fetching token-indexed KV caches, different attention heads are processed with small parallelism on the CSD, thus do not need to be read out at once. In contrast, different token groups are loaded from flash chips at once to compute the attention output within an attention head, which relies on the bandwidth of flash channels and should be read out with maximum parallel.
On the other hand, when writing KV caches to the flash chips, the GPU generates new $k,v$ vectors of all attention heads in parallel. Therefore, enough groups of different attention heads can be batched into flash blocks.
Therefore, for token-indexed KV caches, we prioritize placing groups corresponding to different attention heads within the same block. 
To ensure that each attention head can utilize the full channel bandwidth during load operations, the attention heads are distributed across different blocks and thus stridden across different channels.  Given that the number of groups accessed per read operation during the attention computation is substantially greater than the number of available flash channels (typically 8-16), each channel is fully utilized. %This strategic distribution not only enhances the efficiency of data retrieval and  but also optimizes the overall performance of the system by balancing the load across the available flash channels.

\subsection{Integrate and Scale the System}
\label{sub:scale}
\noindent \textbf{GPU-CSD coordination.} 
In InstInfer, the CSD manages the KV cache created by the GPU during the QKV projection process in both prefilling and decoding phases, and executes the attention computations during decoding. This leads to a pipelined cooperation between the GPU and CSD: 
In the prefilling phase, all computations are handled by the GPU, and the substantial KV cache for all input tokens is transferred to the CSD via the PCIe bus, a process that may be time-consuming. To mitigate this, we implement a layer-wise pipeline wherein the KV caches generated at the $i$-th layer are transferred to the CSD concurrently with the computation at the $(i+1)$-th layer.
During the decoding phase, the CSD performs only attention computations, receiving only the $q,k,v$ vectors from the GPU. After completing these computations, the CSD sends the attention outputs back to the GPU to proceed with generating the $o$ vector and completing the subsequent FFN layer inferences. Compared with the traditional KV cache offloading system, the data volume transmitted on the PCIe buses is reduced by $s/2$, where $s$ refers to the sequence length.
The tasks in the decoding phase between GPU and CSD are executed in an overlapped, mini-batch manner, similar to techniques described in prior works \cite{heo2024neupims}.

For data transmission between the GPU and the CSD, we use a peer-to-peer approach, bypassing the host memory buffer to enable direct data transfer through PCIe lanes. This approach minimizes redundant data copies and optimizes transmission efficiency. Unlike the traditional GPUDirect Storage \cite{GPUDirectRDMA2017} approach, which depends on the host filesystem to manage SSD data, InstInfer operates independently of complex host file systems for managing KV cache. The FTL in InstCSD handles all data mapping and address translation (cf. Section \ref{sub:manage}), storing metadata on its internal DRAM. This configuration significantly reduces host system overhead, as demonstrated in Section \ref{sub:throughput}.

\noindent \textbf{Scale To CSD Array.} 
InstInfer can be seamlessly scaled across multiple CSDs to enhance inference performance significantly. Specifically, mainstream LLMs typically incorporate multi-head attention mechanisms, where each head in a multi-head attention layer computes an independent set of attention scores. 
As each InstCSD in InstInfer exclusively handles the attention module and different attention heads compute independently without inter-dependencies, it becomes feasible to distribute various attention heads across multiple CSDs. For a configuration with $n$ CSDs and $n_{\text{head}}$ attention heads, where typically $n_{\text{head}} \gg n$ (for example, OPT-13b features 40 heads), each CSD can process $n_{\text{head}}/n$ attention heads. Ultimately, the outputs from the attention heads processed on different CSDs are transmitted back to the GPU, which then concatenates these results to form the final output.

\section{Implementation}
\label{sec:impl}
\subsection{System Deployment}
\begin{figure}
    \centering
    \includegraphics[width=0.95\linewidth]{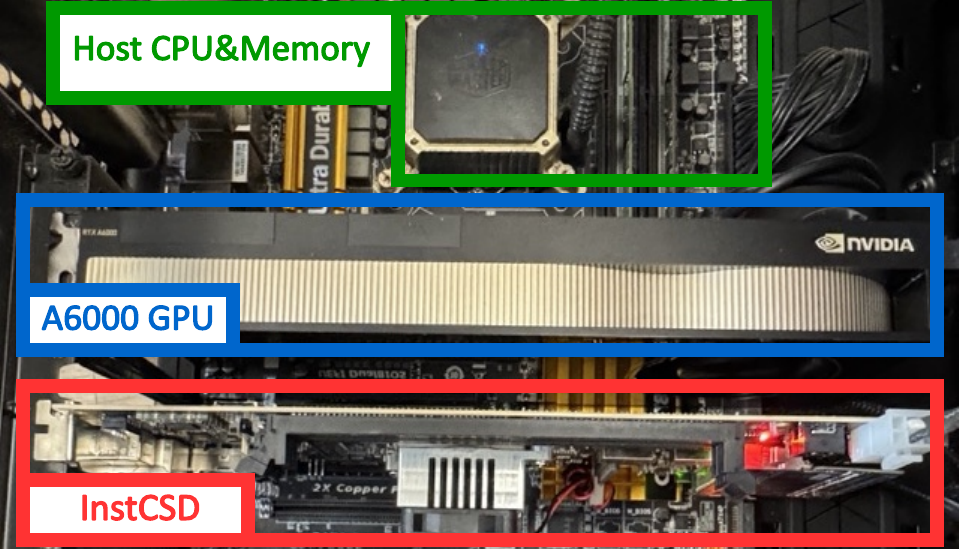}
    \vspace{-5pt}
    \caption{Hardware deployment of InstInfer.}
    \label{fig:impl}
    \vspace{-20pt}
\end{figure}

We have implemented and deployed InstInfer in a real hardware environment, as illustrated in Figure \ref{fig:impl} with comprehensive full-stack software support. InstCSD is built on a Daisyplus OpenSSD, the latest version of a representative CSD device in the OpenSSD project \cite{kwak2020cosmos+, daisplus}. It employs a Xilinx ZU17EG UltraScale+ MPSoC as its processor, which contains a mid-range FPGA chip with a four-core ARM processor and 2GB DRAM. The device is connected to the host through PCIe Gen3x4 lanes as an NVMe drive. 
The hardware SparF Attention engine and NFC filters are implemented on the FPGA part, clocked at a frequency of 285MHz, while the FTL runs as software on the ARM processor. 
The software stack of InstInfer is built on FlexGen \cite{sheng2023flexgen}, which allows us to offload KV cache to the CSD, while maintaining the model weight and activitions on the GPU VRAM. We reconstruct the data path between the GPU and CSD to support peer-to-peer DMA and layer-wise KV cache transmission. The driver for InstCSD is adapted from \cite{Markussen2021smartio}, with specific modifications to NVMe commands to accommodate the unique computational capabilities of InstCSD.

\subsection{Towards Practical CSD Solutions}
While the Daisyplus OpenSSD platform provides an effective environment for deploying a real CSD, it presents several challenges that hinder its practicality for widespread adoption. Notably, the platform features expensive FPGA chips, costing thousands of dollars \cite{daisplus}, and is equipped with a limited storage capacity of only 64GB and merely 4 flash channels. In addition, OpenSSD only supports legacy motherboards like Z97, which lags far behind the current hardware environment \cite{kwak2020cosmos+}. These specifications fall short of the requirements of contemporary SSDs, which typically offer cheaper processors, more channels, and greater storage capacity.

To bridge this gap between experimental setups and practical, cost-effective systems, we adopt NVMeVirt \cite{kim2023nvmevirt}, a cutting-edge software-defined virtual NVMe device. NVMeVirt facilitates a seamless integration with the host software stack like a real NVMe SSD, while providing the flexibility to customize SSD internals with specific needs. 
Therefore, we first collected fine-grained latency and overhead statistics of OpenSSD-based InstCSD deployment with the system and then built the software-defined InstCSD based on the NVMeVirt. To reflect the real computing capability of CSD, we migrated the InstCSD implementation to the Xilinx Zynq7045 \cite{zynq}, a more economically viable FPGA with an SoC that is prevalently utilized in edge computing scenarios.
We further extend the flash channel number to 8 and channel bandwidth to 1.4GB/s to align with modern SSD configurations (i.e., Samesung 980pro \cite{980pro}). 
The detailed resource utilization rates of InstCSD on Zynq7045 are listed in Table \ref{tab:resource}. We exploit the DSP resources of Zynq7045 to deliver the maximum performance for attention computation.

\begin{table}[]
\begin{tabular}{|l|l|l|l|l|}
\hline
\textbf{}            & \textbf{LUT(K)} & \textbf{FF(K)} & \textbf{BRAM Tile} & \textbf{DSP} \\ \hline
Attention Kernel     & 99.2            & 207.3          & 96                 & 768          \\ \hline
Argtopk              & 5.83            & 3.87           & 24                 & 0            \\ \hline
NFC                  & 58.332          & 27.8           & 96                 & 0            \\ \hline
NVMe Controller      & 7.99            & 12.45          & 27.5               & 0            \\ \hline
Interconnect         & 4.12            & 6.17           & 7.5                & 0            \\ \hline
Available            & 218.6           & 437.2          & 545                & 900          \\ \hline
\textbf{Percent(\%)} & 80.27\%         & 58.92\%        & 46.06\%            & 85.33\%      \\ \hline
\end{tabular}
\caption{Resource utilization of InstCSD on Zynq7045.}
\label{tab:resource}
\vspace{-15pt}
\end{table}

\section{Evaluation}
\label{sec:eval}
\subsection{Methodology}
\noindent \textbf{Inference Systems Setup.} We set five LLM inference systems to thoroughly evaluate the performance of InstInfer over the current KV cache offloading systems:

\begin{enumerate}
    \item \texttt{DeepSpeed}: the DeepSpeed-MII system \cite{holmes2024deepspeed} with Zero-Inference \cite{aminabadi2022deepspeed} enabled. It represents the latest KV cache offloading system but supports offloading KV cache only to host memory. We evaluate it to reflect the memory-based offloading scheme.
    \item \texttt{FlexGen}: the FlexGen system \cite{sheng2023flexgen}, which represents the latest KV cache offloading system to both host memory and SSD for throughput-oriented scenarios. We configure its offload target to SSD to evaluate the SSD-based offloading scheme.
    \item \texttt{FlexGen-SparQ}: the \texttt{FlexGen} with SparQ Attention for sparsity with 1/8 compression ratio;
    \item \texttt{InstI-Dense}: our baseline InstInfer implementation without the SparF Attention mechanism;
    \item \texttt{InstI-SparF}: complete InstInfer with SparF Attention for sparsity with 1/8 compression ratio;
\end{enumerate}

\begin{figure*}
\vspace{-8pt}
\centering
\subfloat[Accuracy on the OPT-13b model.]{\label{data:accuracy1}\rotatebox{0}{\includegraphics[width=0.49\linewidth]{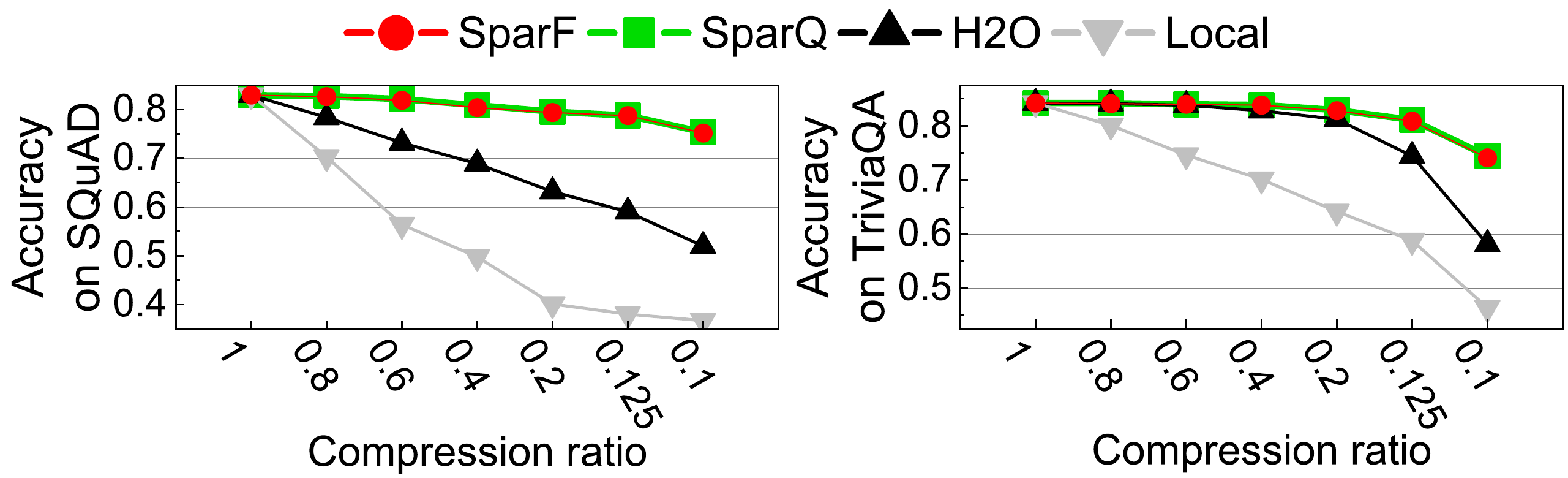}}}
\subfloat[Accuracy on the LLaMA-2-7b model.]{\label{data:accuracy2}\rotatebox{0}{\includegraphics[width=0.49\linewidth]{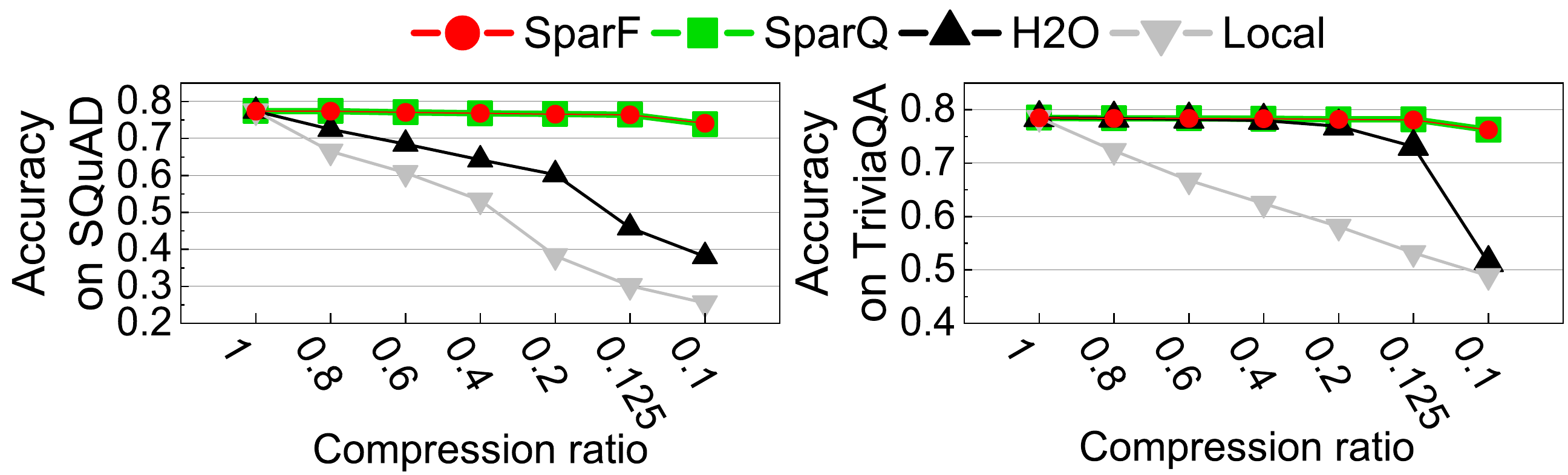}}}
\vspace{-5pt}
\caption{Accuracy of different sparsity methods.}
\vspace{-20pt}
\end{figure*}

\noindent  \textbf{Testbed Configuration.} We conduct our experiments in single CPU-GPU systems. We use NVIDIA A6000 GPU with 48GB VRAM, the 2.2GHz Intel Xeon 5320 CPU with 96GB DDR4 memory, and Samsung 980pro SSDs with 2TB storage. The GPU is connected to CPU via PCIe Gen4x16 lanes. 

\noindent \textbf{Model and Datasets.} We evaluate the OPT-13B model, a representative mid-sized LLM for resource-constraint scenarios. We use FP16 for all variables. The sequences for inference are sampled from popular datasets (i.e., ShareGPT \cite{shareGPT}, Wiki-Text-2 \cite{wiki-text-2}, SQuAD \cite{squad}, and TriviaQA \cite{JoshiTriviaQA2017}).
Both the input and output sequence lengths are set to 1024, matching the maximal context length of OPT-13B to fully demonstrate the long-context scenarios with heavy KV cache burden.

\subsection{Accuracy}
We evaluate the accuracy of the SparF Attention mechanism, along with the vanilla SparQ Attention, H2O \cite{zhang2024h2o}, and the local attention method, under different KV cache compression ratios, as illustrated in Figure \ref{data:accuracy1} and \ref{data:accuracy2}. We observe that our flash-aware SparF Attention performs nearly identically with the vanilla SparQ Attention, while maintaining robustness against the H2O and local attention approaches. This is because SparF primarily focuses on optimizing the KV cache access patterns on the flash chips, efficiently identifying the sparsity in the KV caches with our dual-step loading.
As SparF suffers negligible accuracy loss with a compression ratio up to 1/8, we set the default compression ratio as 1/8 for the following evaluations except particularly noted.

\begin{figure}
    \centering
    \includegraphics[width=0.99\linewidth]{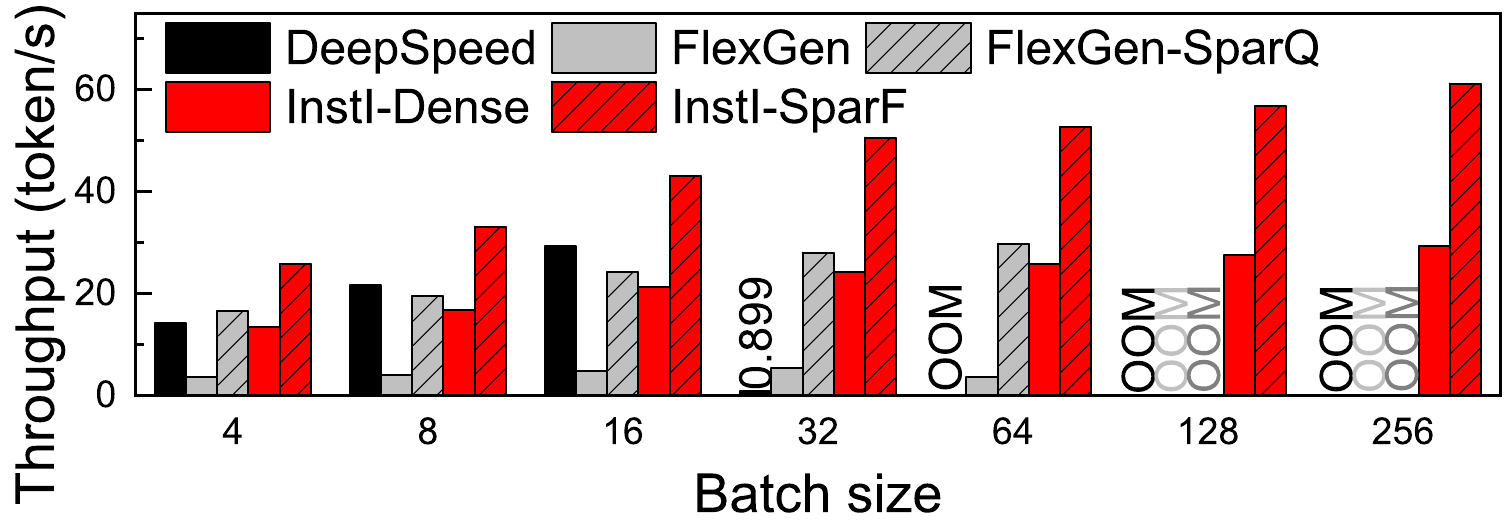}
    \vspace{-15pt}
    \caption{Throughput of different LLM systems: 1-SSD.}
    \label{data:throughput}
    \vspace{-15pt}
\end{figure} 

\subsection{Throughput Evaluation}
\label{sub:throughput}
\noindent \textbf{Performance with 1 SSD (CSD).}
Figure \ref{data:throughput} illustrates the end-to-end throughput of different LLM inference systems with 1 SSD(CSD). The \texttt{DeepSpeed} leverages host memory for KV cache offloading, which owns a larger bandwidth and thereby outperforms other dense schemes when batch size is small (4-16). However, it quickly exceeds the available host memory at bs=32 and incurs the kernel swapping to SSDs, which results in a 32.6$\times$ throughput degradation compared with bs=16. 
\texttt{FlexGen} supports up to bs=64 but delivers much lower throughput due to the SSD-offloading with larger capacity but limited PCIe bandwidth. Note that the OOM error occurs at bs=128 despite the substantial SSD capacity. This is because the intermediate KV cache during the prefilling phase exceeds the available GPU VRAM. In contrast, \texttt{InstI} leverages a layerwise transmission of the KV cache during the prefilling phase, which significantly reduces the VRAM buffer requirement for the intermediate KV caches. 
Therefore, \texttt{InstI} supports much larger batch sizes, and addresses the bandwidth challenges through in-storage attention offloading to significantly mitigate the KV cache transmission overheads in traditional offloading systems, thereby outperforming \texttt{FlexGen} by 6.85$\times$ at bs=64.
\texttt{InstI} shows the best scalability as batch size increases. Note that \texttt{InstI} only outperforms the maximal achievable throughput of the baseline (\texttt{DeepSpeed} at bs=16) by 4.6\% (at bs=256), because the CSD internal bandwidth (11.2GB/s) is still lower than the PCIe bandwidth between GPU and hose memory (32GB/s).
\texttt{InstI-SparF} effectively reduces the demanding KV cache volume, which further improves the throughput of original \texttt{InstI} by up to 2.08$\times$ at bs=256, outperforming the baseline \texttt{FlexGen} by up to 11.1$\times$. This demonstrates the efficiency of SparF Attention for KV cache on the flash chips.

\begin{figure}
    \centering
    \includegraphics[width=0.99\linewidth]{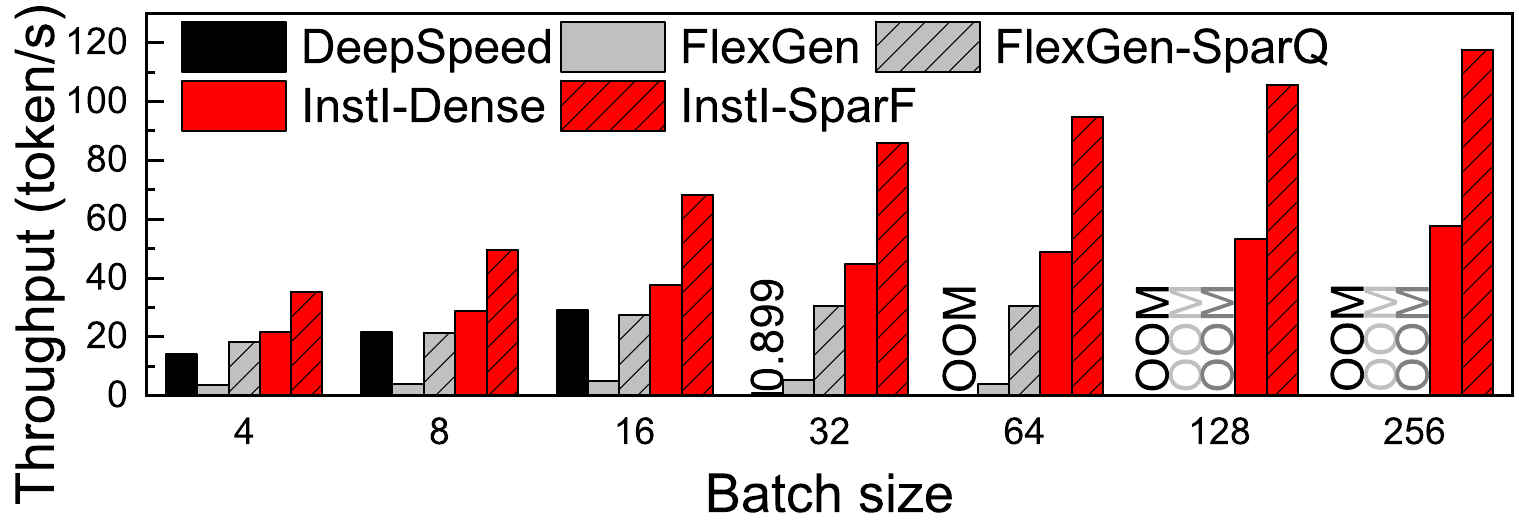}
    \vspace{-15pt}
    \caption{Throughput of different LLM systems: 2-SSD.}
    \label{data:throughput_2}
    \vspace{-15pt}
\end{figure} 

\noindent \textbf{Performance with 2 SSDs (CSDs).}
Figure \ref{data:throughput_2} further presents the throughput with 2 SSDs(CSDs). We observe that traditional offloading schemes exhibit negligible performance improvement despite larger PCIe bandwidth aggregated by multiple SSDs. This is because these schemes rely on the host filesystem to manage KV cache on the SSD, which puts a heavy burden on the data transmission between GPU and SSD. \texttt{InstI} addresses this issue through two approaches. On the one hand, the optimized peer-to-peer DMA transmission between GPU and CSDs bypasses the host; on the other hand, most of the KV cache transmission occurs within the CSD through the internal flash channels, which can be easily scaled up through multiple CSDs. Therefore, \texttt{InstI} (at bs=256) outperform maximal achievable throughput of \texttt{FlexGen} (at bs=32) by 10.5$\times$, and \texttt{InstI-SparF} (at bs=256) outperforms \texttt{FlexGen-SparQ} (at bs=32) by 3.11$\times$, respectively.

\subsection{Latency Breakdown}
\begin{figure}
    \centering
    \includegraphics[width=0.99\linewidth]{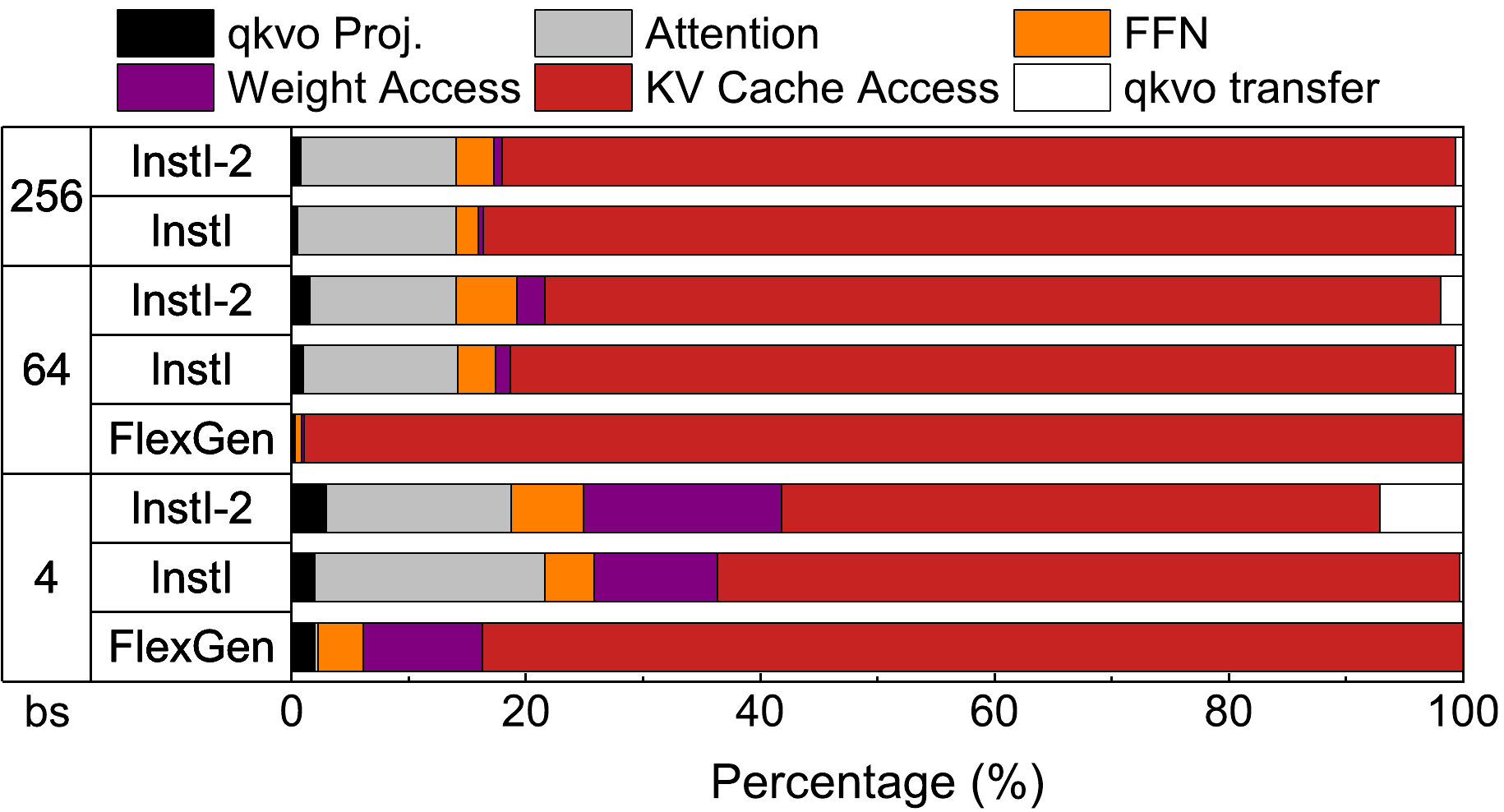}
    \vspace{-15pt}
    \caption{Latency breakdown of dense LLM inference.}
    \label{data:lat_dense}
    \vspace{-10pt}
\end{figure}

\begin{figure}
    \centering
    \includegraphics[width=0.99\linewidth]{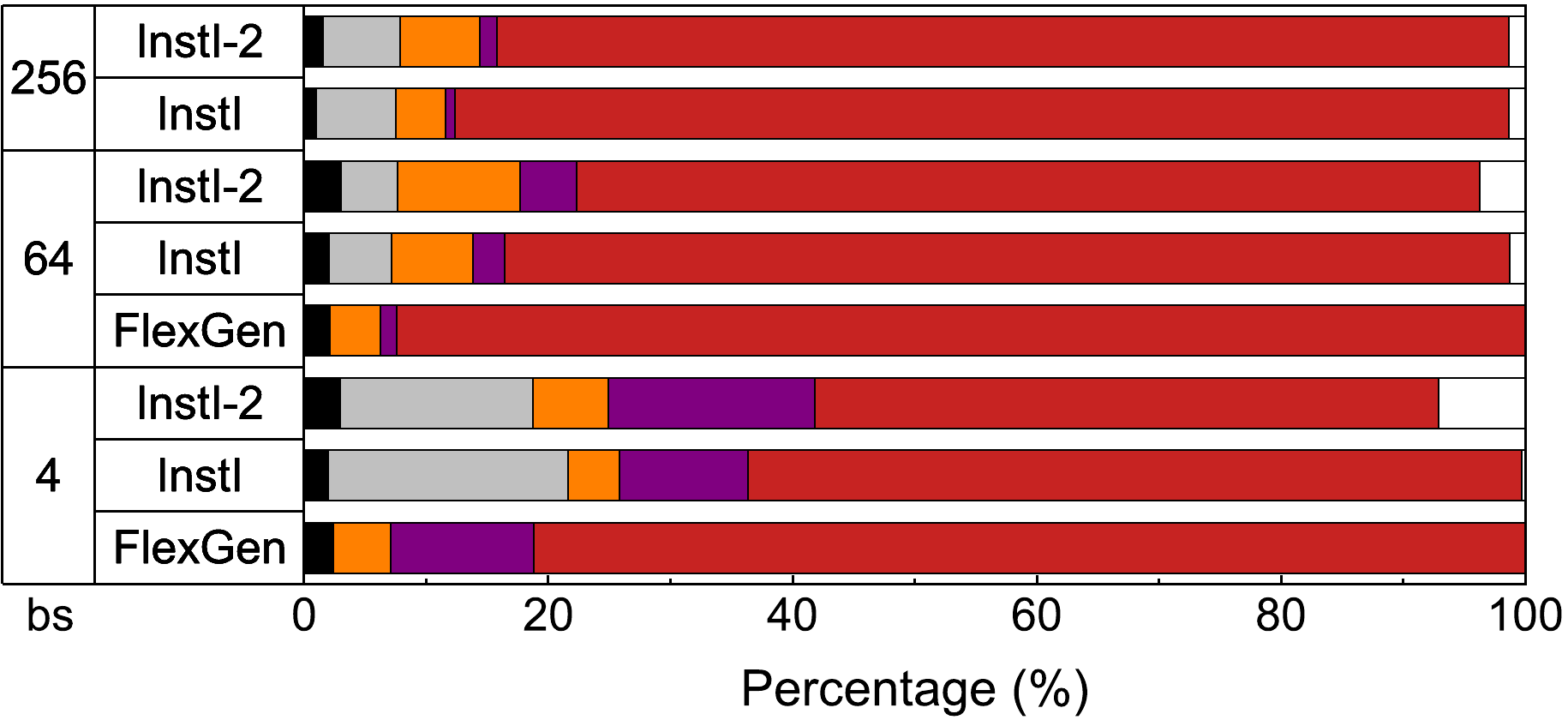}
    \vspace{-15pt}
    \caption{Latency breakdown of sparse LLM inference.}
    \label{data:lat_sparse}
    \vspace{-15pt}
\end{figure}

\noindent \textbf{Decoding Latency Analysis.}
Figure \ref{data:lat_dense} and \ref{data:lat_sparse} depicts the normalized latency breakdown during the decoding phase of different LLM systems, respectively, where \texttt{FlexGen}, \texttt{InstI}, and \texttt{InstI-2} represent FlexGen, InstInfer and InstInfer with 2 CSDs. Both sparse (1/8) and dense attention are evaluated, with small (bs=4), middle (bs=64), and large batch size (bs=256) scenarios. We observe that the KV cache access is the primary bottleneck across all the scenarios and systems, considering the extremely low arithmetic intensity of the attention computation. Nevertheless, compared with \texttt{FlexGen} at bs=64, \texttt{InstI} and \texttt{InstI-2} still reduce the KV cache access percentage from 98.9\% to 80.7\% and 76.4\% in dense inference, and from 92.4\% to 82.3\% and 74.0\%, respectively. In terms of the end-to-end inference latency comparison, the dense \texttt{InstI} and \texttt{InstI-A} reduce the KV cache access overheads by 88.1\% and 94.0\%, respectively.
To further alleviate the KV cache bottleneck, it is promising to scale up to more CSDs or adopt CSD devices with more flash channels based on InstInfer.

\begin{figure}
    \centering
    \includegraphics[width=0.99\linewidth]{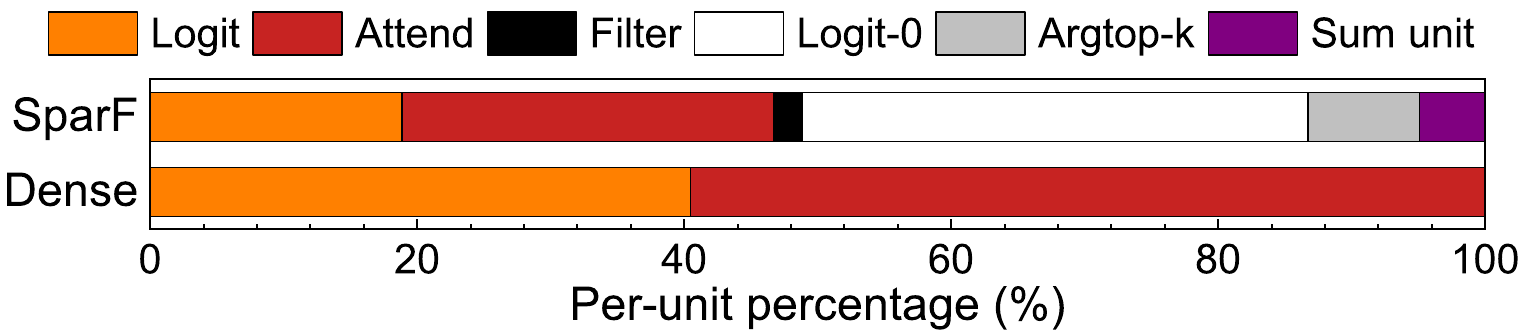}
    \vspace{-15pt}
    \caption{Latency breakdown of the SparF Attention engine.}
    \label{data:engine}
    \vspace{-10pt}
\end{figure}

\noindent \textbf{SparF Attention Engine Analysis.}
We dive into the SparF Attention engine in InstCSD to analyze the normalized overheads of each unit, as illustrated in Figure \ref{data:engine}. Compared with dense attention computation, the primary difference lies in that SparF introduces an additional \texttt{Logit-0} process, corresponding to the step $4$ in Algorithm \ref{algo1}. The extra logit computation further helps identify the sparsity within the sequences, which finally delivers the overall performance improvement.

\subsection{Scalability And Sensitivity Tests}

\begin{figure}
\vspace{-8pt}
\centering
\subfloat[Varying number of InstCSDs.]{\label{data:csds}\rotatebox{0}{\includegraphics[width=0.49\linewidth]{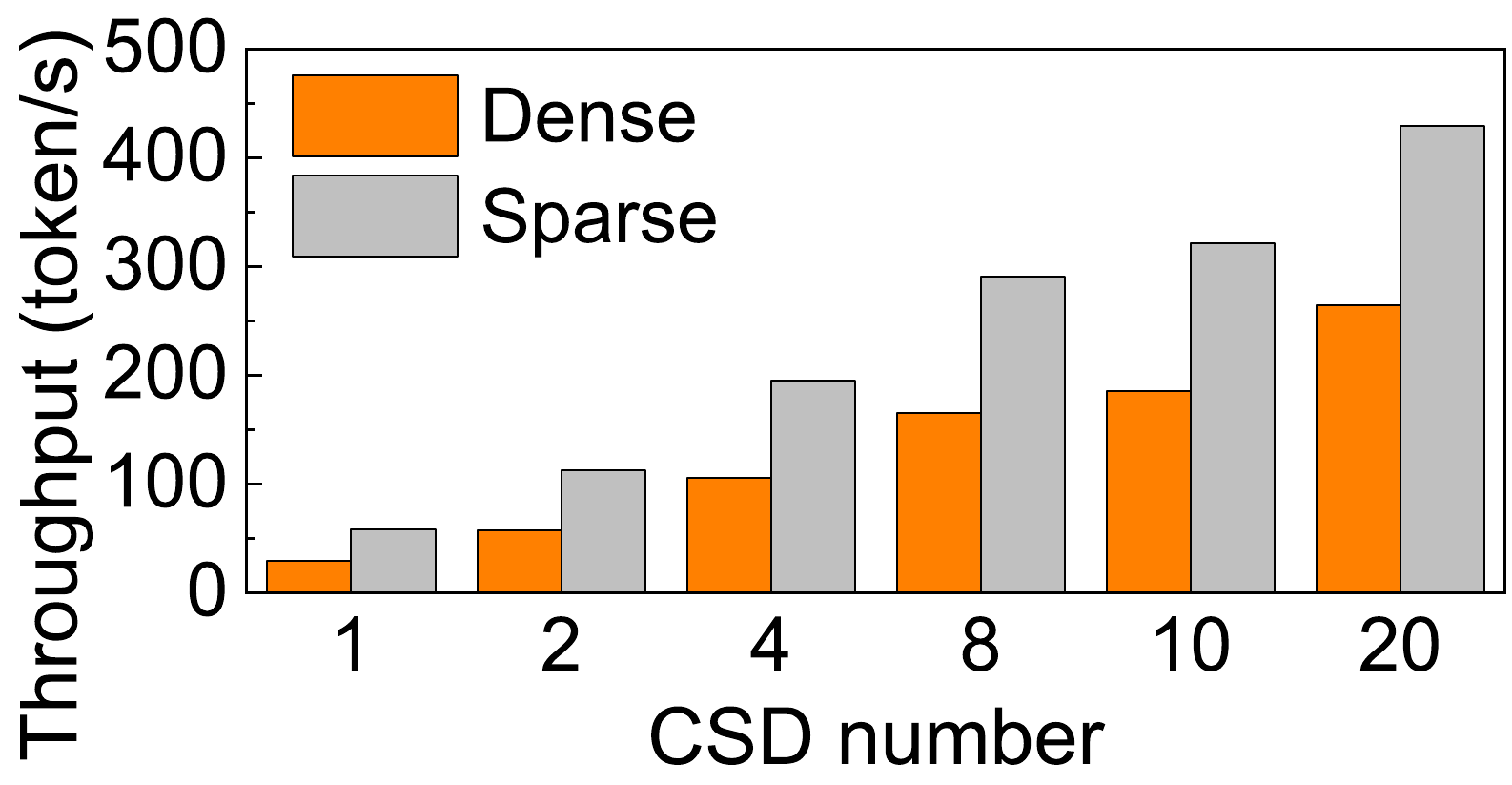}}}
\subfloat[Varying compression ratios.]{\label{data:compression}\rotatebox{0}{\includegraphics[width=0.49\linewidth]{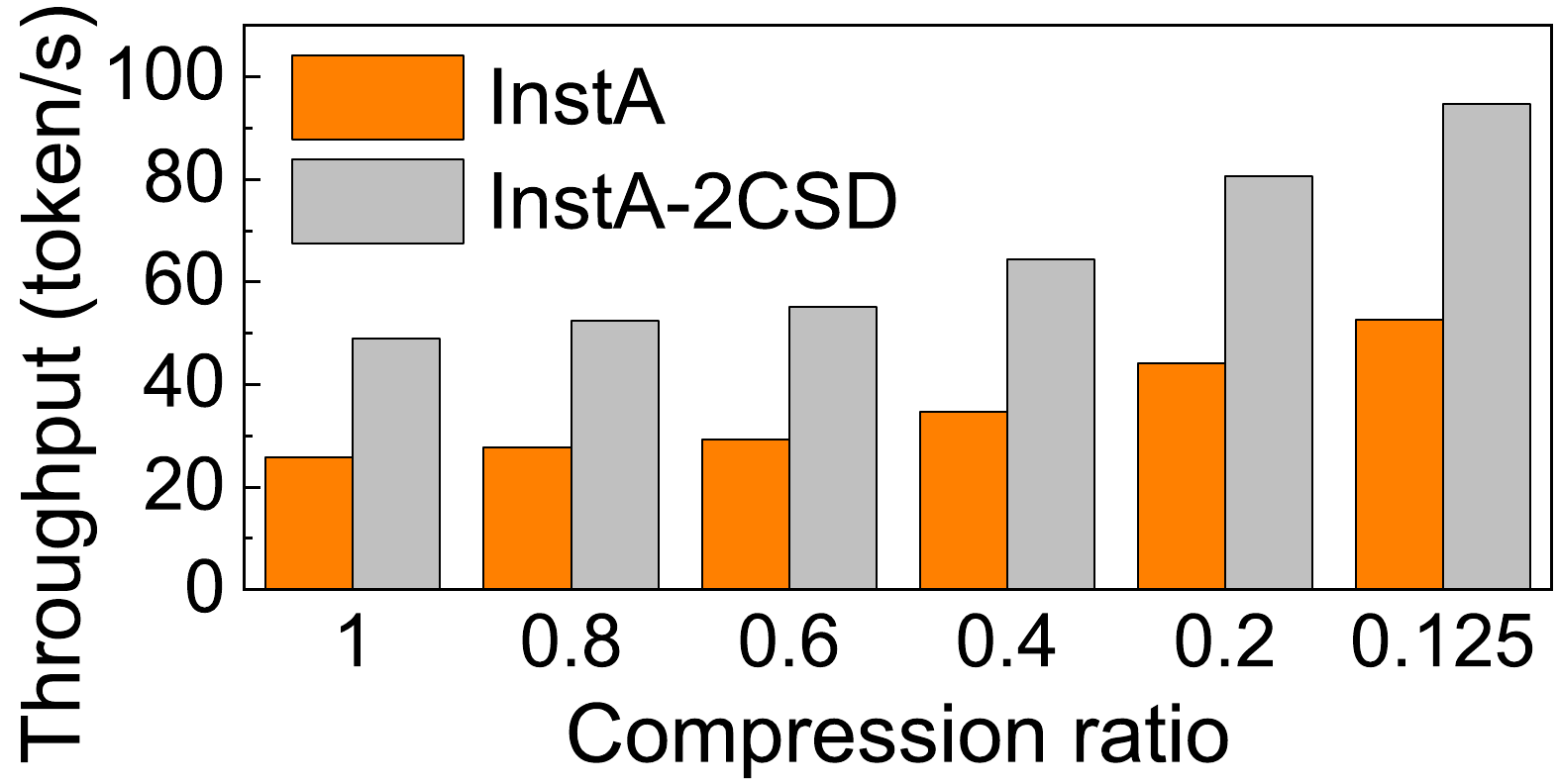}}}
\vspace{-5pt}
\caption{Throughput with varying configurations.}
\vspace{-15pt}
\end{figure}

\noindent \textbf{Scalability with More CSDs.}
We further evaluate the scalability of InstInfer with more CSDs in terms of both dense inference and 1/8-sparsity at bs=256, as depicted in Figure \ref{data:csds} respectively. 
Since traditional KV cache-offloading systems with SSDs show negligible performance improvements scaling with SSD number, we omit their results in the Figure. 
Compared with 1-CSD configuration, deploying 20 CSDs can improve the dense and sparse inference throughput by 8.99$\times$ and 7.29$\times$, respectively. 
The attention head-level parallelism is employed among multiple InstCSDs, which is suitable for InstInfer because only the critical computation (attention operators) and data (KV cache) are offloaded to the CSD. As these computations and data are inherently parallel and have no dependency, the scaling up can be quite straightforward by assigning attention heads to multiple CSDs. Therefore, both the dense and sparse (with SparF Attention) \texttt{InstI} shows good scalability with an increasing number of CSDs. 

\noindent \textbf{Sensitivity with Varying Sparsity.}
Figure \ref{data:compression} shows the throughput of \texttt{InstI} with 1 or 2 CSDs under different compression ratios with SparF Attention, respectively. We observe that although a larger compression ratio leads to more random fine-grained access to KV cache on the flash chips, which is typically a challenge for SSDs, InstInfer efficiently benefits from larger compression ratios due to the efficient dual-step loading mechanism of SparF Attention.

\section{Related Works And Discussion}
\label{sec:related}
\noindent \textbf{PIM-Based Transformer Acceleration.}
Several works \cite{heo2024neupims, choi2023unleashing, zhou2022transpim, wu2024pim} leverage Processing-In-Memory (PIM) techniques to address the storage and bandwidth bottleneck of LLM inference, which integrate computing units within the memory cells to accelerate the memory-bound attention computation. However, these works are all based on simulators, considering that PIM devices are still expensive to manufacture and far from being widely deployed in practice, especially for resource-constraint scenarios. Additionally, it only partially addresses the KV cache capacity issues compared with the abundant storage capacity of flash chips. In contrast, InstInfer is deployed in real hardware, adopting economical FPGA chips and SSDs as a more cost-effective and scalable solution to address the storage and bandwidth challenges.

\noindent \textbf{Optimizations For KV Cache Management.}
vLLM \cite{kwon2023efficient} manages KV cache in GPU VRAM and host memory in block-granularity, which takes inspiration from the virtual memory mechanism, to reduce the overhead of fragmentation. LMDeploy \cite{lmdeploy} and CachedAttention \cite{gao2024cachedattention} focus on managing KV caches on the host memory and SSDs to reduce the recomputation overheads in multi-turn conversations. These works aim to optimize the prefilling phase in online inference scenarios, which rely on the scheduling of KV cache across multiple memory tiers. However, these approaches are not suitable for inference with long output sequences, which are primarily constrained by the large volume of KV cache transmission during the decoding phase.
Other solutions \cite{patel2023splitwise, zhong2024distserve, Wonbem2024InfiniGen, qin2024mooncake, he2024fastdecode} leverage disaggregated resources (i.e., GPU, CPU and memory pools) to store the KV cache and accelerate long-context LLM inference, which are not suitable for resource-constraint scenarios. 
InstInfer leverages cost-effective CSDs, which are more applicable and effectively address the decoding-phase bottleneck of KV cache.

\section{Conclusion}
In this work, we introduced InstInfer, a novel CSD-based LLM offline inference system to address the substantial storage and bandwidth challenges associated with KV caches in a cost-effective approach. By offloading the critical decoding-phase attention and KV cache to CSDs with flash-aware designs, InstInfer exploits high channel bandwidth of flash chips, circumventing the limitations imposed by external PCIe bandwidth. Our evaluation shows that InstInfer outperforms current SSD-offloading systems by up to 11.1$\times$ for long-context inference in resource-constraint scenarios.

%%%%%%%%% -- BIB STYLE AND FILE -- %%%%%%%%
\bibliographystyle{IEEEtranS}
\bibliography{main}
% Generated by IEEEtranS.bst, version: 1.13 (2008/09/30)

\end{document}